\documentclass[pra, twocolumn, final, reprint, amsmath, amssymb, superscriptaddress, showpacs, a4paper, aps, 10pt, hidelinks, numbers, sort&compress, floats]{revtex4-1} 

\usepackage{epsfig}
\usepackage{graphicx}
\usepackage[normalem]{ulem}
\usepackage{xcolor}
\usepackage{amsmath, amssymb, amsthm, times, graphics, bm, bbm, dcolumn}
\usepackage[colorlinks,urlcolor=black,citecolor=blue,linkcolor=black]{hyperref}

\newcommand{\previous}[1]{}

\newcommand{\ket}[1]{\mbox{\ensuremath{\vert #1 \rangle}}}

\newcommand{\br}{{\bf r} }

\newcommand{\refeq}[1]{\mbox{Eq.~(\ref{#1})}}

\begin{document}

\title{
Time-resolved observation of competing attractive and repulsive\\short-range correlations in strongly interacting Fermi gases}
\author{A. Amico}
\altaffiliation[These authors contributed equally to this work.]{}
\affiliation{\mbox{LENS and Dipartimento di Fisica e Astronomia, Universit\`{a} di Firenze, 50019 Sesto Fiorentino, Italy}}
\affiliation{Istituto Nazionale di Ottica del Consiglio Nazionale delle Ricerche (INO-CNR), 50019 Sesto Fiorentino, Italy}
\author{F. Scazza}
\altaffiliation[These authors contributed equally to this work.]{}
\affiliation{\mbox{LENS and Dipartimento di Fisica e Astronomia, Universit\`{a} di Firenze, 50019 Sesto Fiorentino, Italy}}
\affiliation{Istituto Nazionale di Ottica del Consiglio Nazionale delle Ricerche (INO-CNR), 50019 Sesto Fiorentino, Italy}
\author{G. Valtolina}
\altaffiliation[Present address: ]{JILA, University of Colorado, Boulder, CO 80309, USA}
\affiliation{\mbox{LENS and Dipartimento di Fisica e Astronomia, Universit\`{a} di Firenze, 50019 Sesto Fiorentino, Italy}}
\affiliation{Istituto Nazionale di Ottica del Consiglio Nazionale delle Ricerche (INO-CNR), 50019 Sesto Fiorentino, Italy}
\author{P. E. S.~Tavares}
\altaffiliation[Present address: ]{Instituto de F\'isica de S\~{a}o Carlos, Universidade de S\~{a}o Paulo, Caixa Postal 369, 13560-970 S\~{a}o Carlos, SP, Brazil}
\affiliation{\mbox{LENS and Dipartimento di Fisica e Astronomia, Universit\`{a} di Firenze, 50019 Sesto Fiorentino, Italy}}
\affiliation{Istituto Nazionale di Ottica del Consiglio Nazionale delle Ricerche (INO-CNR), 50019 Sesto Fiorentino, Italy}
\author{W. Ketterle}
\affiliation{Department of Physics, MIT-Harvard Center for Ultracold Atoms, and Research Laboratory of Electronics, MIT, Cambridge, Massachusetts 02139, USA}
\author{M. Inguscio}
\affiliation{\mbox{LENS and Dipartimento di Fisica e Astronomia, Universit\`{a} di Firenze, 50019 Sesto Fiorentino, Italy}}
\affiliation{Istituto Nazionale di Ottica del Consiglio Nazionale delle Ricerche (INO-CNR), 50019 Sesto Fiorentino, Italy}
\author{G. Roati}
\affiliation{\mbox{LENS and Dipartimento di Fisica e Astronomia, Universit\`{a} di Firenze, 50019 Sesto Fiorentino, Italy}}
\affiliation{Istituto Nazionale di Ottica del Consiglio Nazionale delle Ricerche (INO-CNR), 50019 Sesto Fiorentino, Italy}
\author{M. Zaccanti}
\email[Corresponding author. E-mail: ]{matteo.zaccanti@ino.cnr.it}
\affiliation{\mbox{LENS and Dipartimento di Fisica e Astronomia, Universit\`{a} di Firenze, 50019 Sesto Fiorentino, Italy}}
\affiliation{Istituto Nazionale di Ottica del Consiglio Nazionale delle Ricerche (INO-CNR), 50019 Sesto Fiorentino, Italy}

\begin{abstract}
We exploit a time-resolved pump-probe spectroscopic technique to study the out-of-equilibrium dynamics of an ultracold two-component Fermi gas, selectively quenched to strong repulsion along the upper branch of a broad Feshbach resonance. 
For critical interactions, we find the rapid growth of short-range anti-correlations between repulsive fermions to initially overcome concurrent pairing processes. At longer evolution times, these two competing mechanisms appear to macroscopically coexist in a short-range correlated state of fermions and pairs, unforeseen thus far. 
Our work provides fundamental insights into the fate of a repulsive Fermi gas, and offers new perspectives towards the exploration of complex dynamical regimes of fermionic matter. 
\end{abstract}

\maketitle
\onecolumngrid
\vspace*{-5pt}
\twocolumngrid

\noindent 
Ultracold quantum gases offer a pristine platform for the realization of minimal Hamiltonians, enabling to investigate the relationship between the macroscopic behavior of a many-body system and the interactions between its constituents. In this context, a two-component atomic Fermi gas with tunable short-range repulsive interactions has attracted a growing interest, being regarded as a paradigmatic framework to address the Stoner model of itinerant ferromagnetism \cite{Stoner1933, Duine2005}.  
However, in contrast with the widely explored case of attractively interacting Fermi gases \cite{Varenna2007,Zwerger2012},  
the nature of the repulsive Fermi gas and its instability towards a ferromagnetic state remain largely debated, both in theory \cite{Zhai2009, Pilati2010, Cui2010, Shenoy2011, Zhang2011, Chang2011, Pekker2011, Sodemann2012, Massignan2013, Massignan2014, He2016} and experiments \cite{Jo2009, Sanner2012, Kohstall2012, Trotzky2015, Scazza2017, Valtolina2017}. 
This stems from the fact that genuine short-range repulsion only develops along an excited \emph{upper} energy branch of the many-body system \cite{Shenoy2011, Massignan2014}, which is unstable against relaxation into the paired many-body ground state \cite{Varenna2007, Zwerger2012}, i.e.~the \emph{lower} branch. 
As a consequence, it is still questioned whether a repulsive Fermi liquid exists at strong coupling, whether ferromagnetic correlations can develop therein, and how they possibly compete with pairing correlations.

\begin{figure*}[t]
\begin{center}
\includegraphics[width=178mm]{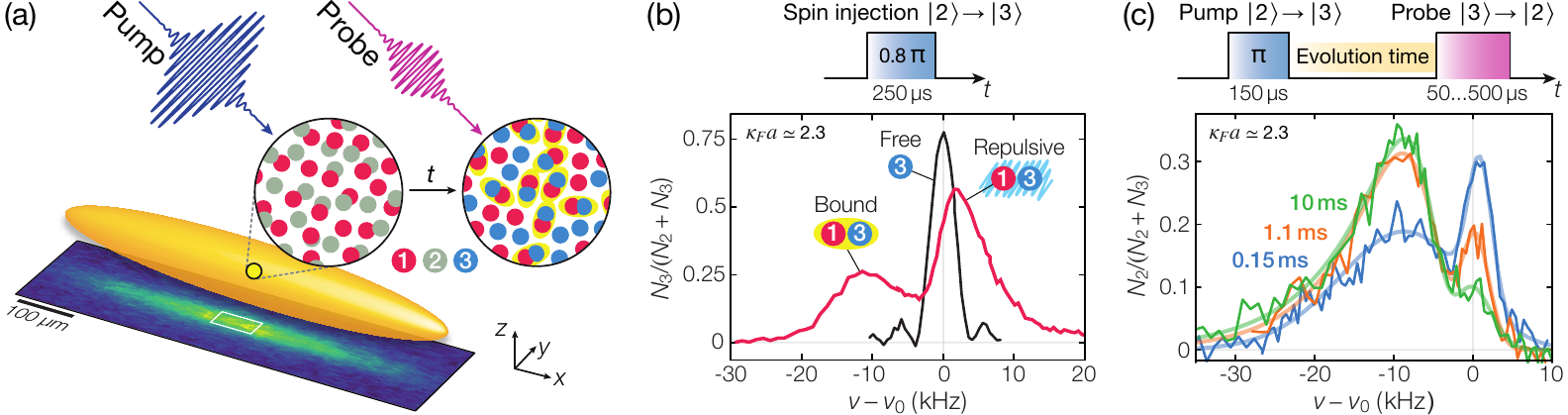}
\caption{
\textbf{(a)} 
A pump pulse %
converts a weakly interacting 1-2 $^6$Li mixture into a 1-3 strongly repulsive one. %
After a variable evolution time the gas is probed by a second pulse. The spectroscopic signal is acquired within a central region of the cloud, denoted by a white rectangle in the image.
\textbf{(b)} 
Spin-injection spectrum at $\kappa_F a \simeq 2.3$ (red line). Both the atomic repulsive resonance located at $\Delta_{+0}>0$ and the pair association spectrum are visible, %
referenced to the spectrum of a spin-polarized state-$2$ Fermi gas (black line). 
The pump pulse remains fully selective up to $\kappa_F a \approx 2.5$, where the atomic and molecular spectral contributions overlap.
\textbf{(c)} 
Probe spectra at $\kappa_F a \simeq 2.3$ for various evolution times (see legend),
fitted to a phenomenological model consisting of the sum of Gaussian and Gumbel functions \cite{SI}. The spin-injection and pump-probe spectroscopy pulse sequences are also sketched.
 }
\label{Fig1}
\end{center}
\vspace*{-0pt}
\end{figure*}

Here, we address these open questions by studying the time-resolved spectral response of a balanced spin mixture of ultracold $^6$Li atoms. 
We first employ spin-injection radio-frequency (RF) spectroscopy to precisely locate the upper branch, %
finding that it remains well defined up to strong couplings. Then, in analogy with ultrafast optical spectroscopy in solid state \cite{Orenstein2012, Giannetti2016}, we exploit a pump-probe RF scheme to investigate the out-of-equilibrium dynamics of a strongly repulsive Fermi liquid. A first \emph{pump} pulse selectively transfers the gas to the upper branch, while a second \emph{probe} spectroscopy pulse monitors the following evolution [see Fig.~\ref{Fig1}(a)] with a resolution of few Fermi times $\tau_F$, which sets the minimum collective response time in fermionic many-body systems \cite{Cetina2016}.
By tracking the evolution of the atomic probe spectra center and amplitude, we observe the build-up of atomic anti-correlations in the upper branch and the onset of pairing processes into the lower branch.
We extract the initial growth rates for both these mechanisms, developing over time scales of few $\tau_F$. In contrast with theoretical predictions for an instantaneously quenched Fermi gas \cite{Pekker2011}, the rate associated with the build-up of short-range correlations between repulsive quasi-particles is found to be faster than the growth rate of pairing processes, albeit comparable with it. 
At longer evolution times, we observe an unpredicted slowly evolving regime, where a minority population of unpaired fermions coexists with pairs in a short-range correlated state. 

We produce weakly interacting mixtures of about $2 \times 10^5$ $^6$Li atoms in a cylindrically-shaped optical dipole trap \cite{Scazza2017, SI} [see Fig.~\ref{Fig1}(a)], equally populating the two lowest hyperfine atomic states, hereafter denoted as 1 and 2 respectively. Unless otherwise specified, the initial system temperature is $T =0.12(2)\,T_F$, where $T_F$ is the Fermi temperature \cite{SI}.
Our spin-injection spectroscopy protocol for balanced mixtures is analogous to that employed for investigating repulsive Fermi polarons \cite{Scazza2017}.  
We typically employ 250\,$\mu$s-long RF pulses, corresponding to a 0.8$\pi$-pulse for non-interacting atoms, whose detuning $\delta \nu = \nu-\nu_0$ from the bare $2 \rightarrow 3$ transition is scanned. Here, 3 denotes the third-to-lowest $^6$Li hyperfine state. 
To adjust the interparticle interaction strength, we exploit a broad Feshbach resonance between states 1 and 3 located at 690\,G \cite{SI}.
For each magnetic field between 640\,G and 680\,G, the spectroscopy signal is defined by the fraction of transferred atoms, recorded within the central region of the cloud to reduce the effects of density inhomogeneity [see Fig.~\ref{Fig1}(a)]. The initial average density $\overline{n}_{0}$ of state-2 atoms within this region sets the relevant Fermi energy $\varepsilon_F$ and wavevector $\kappa_F$ of the gas \cite{Scazza2017,SI}.
The spectral response exhibits two distinct features [see Fig.~\ref{Fig1}(b)]: an incoherent contribution at $\delta \nu <0$, related to the binding of two fermions into a 1-3 molecule, clearly detectable only at strong interactions, and a coherent atomic resonance at $\delta \nu > 0$. The latter is associated with the conversion of a weakly interacting mixture into a strongly repulsive Fermi liquid.
The center position $\Delta_{+0}$ of this \emph{atomic} peak encodes information about the energy and effective mass of repulsive quasi-particles and their mutual interactions \cite{Massignan2014, Scazza2017, Varenna2014}.

We then exploit the knowledge of $\Delta_{+0}$ to selectively quench weakly interacting mixtures into the strongly repulsive regime.
For each magnetic-field value, an optimized RF pump pulse of about 150\,$\mu\textrm{s} \sim 6\,\tau_F$, with $\tau_F=\hbar/\varepsilon_F \approx 25\,\mu$s, enables a coherent $2 \rightarrow 3$ population transfer to the 1-3 upper branch with high efficiency, exceeding 70$\%$ for the strongest couplings investigated here \cite{SI}. Immediately after the pump pulse, a 3\,$\mu$s-long optical blast removes leftover state-2 atoms while negligibly affecting the other spin components. 
The out-of-equilibrium dynamics of the 1-3 mixture at different interaction strengths is monitored by performing RF spectroscopy on the $3\rightarrow2$ transition, using a probe pulse applied at variable hold time after the quench.
Examples of time-resolved probe spectra are shown in Fig.~\ref{Fig1}(c). 
From these, we characterize the evolution of the atomic peak at $\delta \nu > 0$ by extracting its amplitude $A(t)$ and center $\Delta_+(t)$ through Gaussian fits.
Owing to their halo character, the weakly bound 1-3 dimers created via inelastic collisions remain optically trapped, and atom-to-molecule conversion does not cause any detectable particle loss. Hence, the decay of $A(t)$ directly reflects the drop of the average fermion density $\overline{n}(t)$ and quantifies the development of pairing processes.
Conversely, the peak center $\Delta_+(t)$ provides a measure of the instantaneous (column-integrated) average interaction energy experienced by the surviving state-3 fermions due to the surrounding state-1 atoms and 1-3 pairs.
For contact interactions, this is directly linked to the local pair correlator at zero distance $\langle \psi_1^\dagger(\textbf{r}) \psi_3^\dagger(\textbf{r}) \psi_3(\textbf{r}) \psi_1(\textbf{r}) \rangle$, $\psi_{\sigma}$ being state-$\sigma$ fermion annihilation operator \cite{Varenna2014}. $\Delta_+(t)$ is also proportional to the contact $C$ \cite{Varenna2014}, that quantifies the number of repulsive fermion pairs at short distance also in out-of-equilibrium systems \cite{Bardon2014, Fletcher2016}. Therefore, $\Delta_+(t)$ is sensitive to the development of short-range anti-correlations in the upper branch and to spin-polarized domain formation, both causing a substantial drop of the pair correlator.

\begin{figure}[t!]
\begin{center}
\includegraphics[width=86mm]{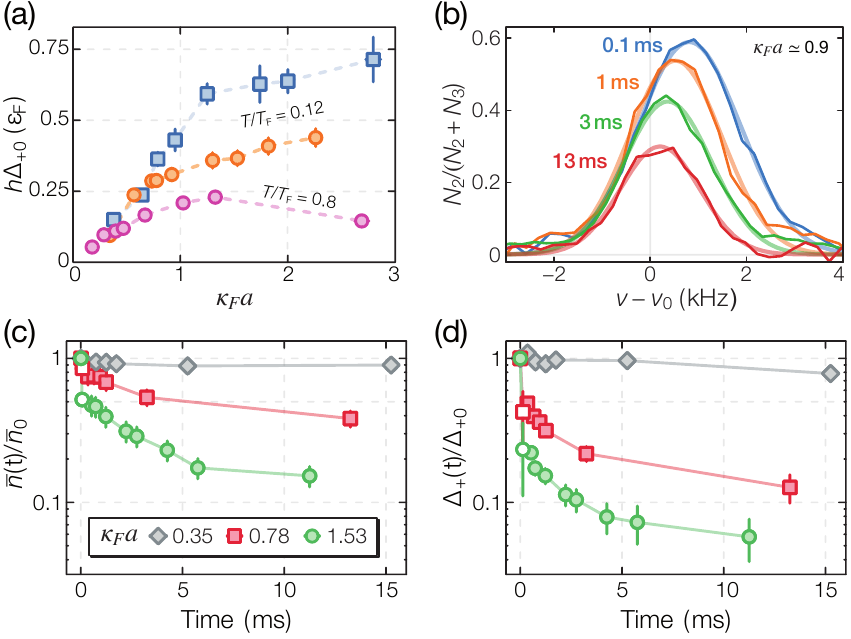}
\caption{
\textbf{(a)} 
Interaction shift $h \Delta_{+0}/\varepsilon_F$ as a function of $\kappa_F a$, measured in balanced mixtures at $T/T_F \simeq 0.12$ (orange circles) and $T/T_F \simeq 0.8$ (magenta circles), and in the impurity limit (blue squares) \cite{Scazza2017}.
Vertical error bars combine the standard error of the Gaussian fits to the atomic spectra with the uncertainty in determining $\varepsilon_F$. Horizontal error bars account for the uncertainty in determining $\kappa_F$.
\textbf{(b)} Examples of atomic probe spectra at $\kappa_F a \simeq 0.9$ for various times after the quench (see legend), together with Gaussian fits.
\textbf{(c)}-\textbf{(d)} Evolution of relative fermion density $\overline{n}(t)/\overline{n}_0$ and shift $\Delta_+(t)/\Delta_{+0}$ for different $\kappa_F a$ values (see legend). 
Colored (white) data points are obtained with a 250\,$\mu$s-long (50\,$\mu$s-long) probe pulse.
Error bars account for the standard error of the Gaussian fit, and in panel \textbf{(c)} also for the uncertainty in the calibration of $\overline{n}_0$ \cite{SI}.
}
\label{Fig2}
\end{center}
\vspace*{-5pt}
\end{figure}

Figure \ref{Fig2}(a) summarizes the results of our spin-injection spectroscopy studies, shown for $T/T_F \simeq 0.12$ and 0.8. The measured $\Delta_{+0}$, normalized to the Fermi energy $\varepsilon_F$, is displayed as a function of the 1-3 interaction strength, parametrized by $\kappa_F a$, where $a\equiv a_{13}$ is the $s$-wave scattering length. 
For both low (orange circles) and high (magenta circles) temperatures, $\Delta_{+0}$ is found to progressively increase with increasing repulsion, featuring a trend qualitatively analogous to that observed in the impurity limit of a spin-imbalanced gas (blue squares) \cite{Scazza2017}.
As for this latter case, we attribute the saturation of $\Delta_{+0}$ at strong coupling to mass renormalization of state-2 fermions converted into heavier repulsive quasi-particles \cite{SI}.
Our data demonstrates the existence of a well-defined repulsive branch for $\kappa_F a \lesssim 2.5$, even in the regime of moderate degeneracy, in spite of an increased atomic spectral width due to enhanced momentum relaxation expected within Fermi liquid theory \cite{SI, Cetina2015}.

Figure~\ref{Fig2}(b) displays typical time-resolved atomic probe spectra, acquired at $\kappa_F a \simeq 0.9$, together with Gaussian fits used to extract $A(t)$ and $\Delta_+(t)$. Examples of the obtained evolutions at various interaction strengths are shown in Fig.~\ref{Fig2}(c)-(d), where both quantities have been normalized to their initial values \cite{SI}.
In the absence of sizeable particle losses, the trend of $A(t)/A(0)$ is equivalent to that of the relative upper-branch population $\overline{n}(t)/\overline{n}_{0}$ [see Fig.~\ref{Fig2}(c)]. 
While for weak repulsion $\overline{n}(t)/\overline{n}_{0}$ and $\Delta_+(t)/\Delta_{+0}$ show only small and slow variations, their dynamics change qualitatively once the gas is quenched to $\kappa_F a\gtrsim 0.8$. There, both observables exhibit a considerable reduction over sub-millisecond timescales, progressively faster and more pronounced for increasing $\kappa_F a$. This strongly points to the onset of the pairing instability and to the concurrent growth of anti-correlations between repulsive quasi-particles, possibly arising from a competing ferromagnetic instability~\cite{Pekker2011, Valtolina2017}. 

\begin{figure}[b!]
\begin{center}
\includegraphics[width=86mm]{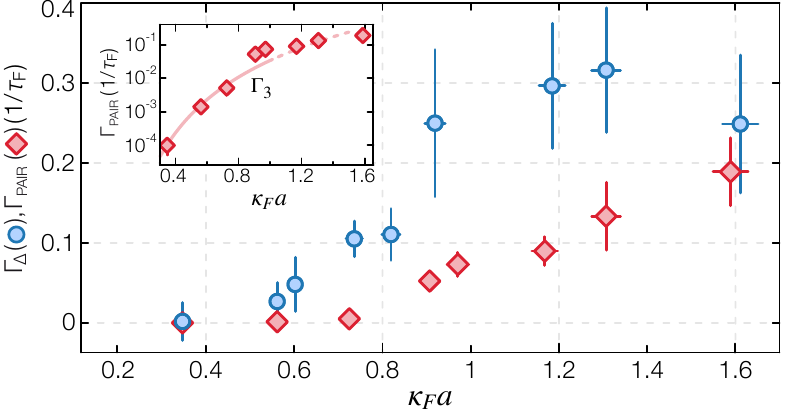}
\caption{Growth rates of anti-correlations in the upper branch $\Gamma_{\Delta}$ (blue circles) and of pairing in the lower branch $\Gamma_\textrm{pair}$ (red diamonds), extracted through linear fits to the measured short-time dynamics of $\Delta_+(t)/\Delta_{+0}$ and of $\overline{n}(t)/ \overline{n}_{0}$, respectively. 
Vertical error bars combine the standard error of the fits with the uncertainty in determining $\tau_F$. 
Horizontal error bars account for the uncertainty in the determination of $\kappa_F$. 
Inset: $\Gamma_{\text{pair}}$ is compared with the predicted three-body recombination rate $\Gamma_3$ \cite{SI, Petrov2003}.
}
\label{Fig3}
\end{center}
\end{figure}

We gain quantitative insights into these competing processes by extracting the initial growth rates $\Gamma_{\text{pair}}$ and $\Gamma_{\Delta}$. To this end, we use linear fits to the evolution at $t \leq 200\,\mu$s of $\overline{n}(t)/\overline{n}_{0}$ and of $\Delta_+(t)/\Delta_{+0}$, respectively; the linear fits represent the time derivatives of $\overline{n}(t)$ and $\Delta_+(t)$. To increase the time resolution of our spectroscopy protocol, we employ probe pulses as short as 50$\,\mu$s, yielding  
for a minimum hold time of $10\,\mu$s a highest measurable rate $\Gamma_{\text{max}} \sim 0.4\,\tau_F^{-1}$. When approaching $\Gamma_{\text{max}}$, the measured rates should be regarded as lower bounds of  the actual ones. For $\kappa_F a \geq 0.8$ both rates increase considerably, with $\Gamma_{\Delta}$ larger than $\Gamma_{\text{pair}}$ (see Fig.~\ref{Fig3}). In particular, whereas the former saturates already for $\kappa_F a \sim 1$ near $\Gamma_{\text{max}}$, $\Gamma_{\text{pair}} \simeq 0.2\,\tau_F^{-1}$ at $\kappa_F a \simeq 1.6$.  
In contrast with theoretical expectations \cite{Pekker2011}, we find that atom-atom correlations in the upper branch grow initially faster than the pairing ones. Differently from recent studies of collective spin dynamics \cite{Valtolina2017}, the present measurements can identify emerging repulsive short-range correlations, but cannot discern whether these develop within a paramagnetic Fermi liquid \cite{Zhai2009}, or herald instead a ferromagnetic instability.
Nonetheless, $\Gamma_{\Delta}$ reasonably agrees with the predicted growth rate of the most unstable mode of the ferromagnetic instability for a zero-temperature homogeneous Fermi gas \cite{Pekker2011}. 
This is expected to establish short-range ferromagnetic correlations at wavevectors of order $\kappa_F/2$ for $\kappa_F a \gtrsim 1$ \cite{Pekker2011, Sodemann2012, Sanner2012}, promoting the development of ferromagnetic order over a length $\xi \sim (\pi/\kappa_F)(\Gamma_{\Delta} \tau_F)^{-1/2} \approx 2\pi/\kappa_F \simeq 2.5\,\mu$m, or $1.6$ interparticle spacings.
Conversely, $\Gamma_{\text{pair}}$ is about one order of magnitude lower than the pairing rate obtained in Ref.~\cite{Pekker2011}, whereas it fits well to the rate $\Gamma_3$ expected for three-body recombination processes \cite{Petrov2003, SI} (see inset of Fig.~\ref{Fig3}). While this result matches that obtained in the impurity limit \cite{Scazza2017}, our data also agree with previous measurements in balanced mixtures \cite{Sanner2012}. One reason for the sizable theory-experiment mismatch may be found in the spectral selectivity of the quench protocol employed in this work, and presumably also in Ref.~\cite{Sanner2012}. This starkly contrasts with the instantaneous quench considered in Ref.~\cite{Pekker2011}, which projects a non-interacting Fermi gas onto all available (atomic and molecular) many-body states of the resonant mixture.

\begin{figure*}
\begin{center}
\includegraphics[width=178mm]{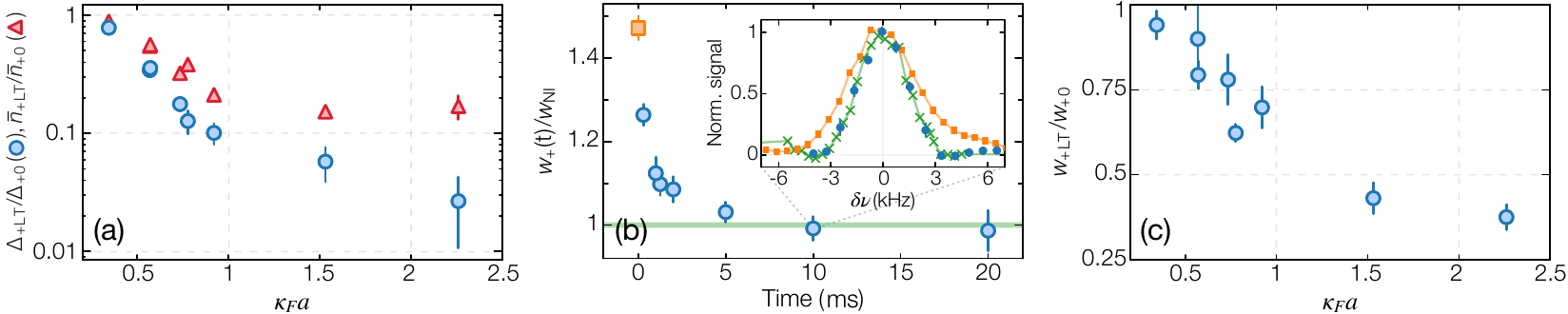}
\caption{
\textbf{(a)} Relative atomic populations $\overline{n}_{LT}/\overline{n}_{0}$ (red triangles) and interaction shifts $\Delta_{+LT}/\Delta_{+0}$ (blue circles), obtained by averaging data points at evolution times $10 \leq t \leq 30$ ms, as a function of $\kappa_F a$. 
\textbf{(b)} Evolution of the atomic peak full width at half maximum $w_{+}(t)$ at $\kappa_F a \simeq 0.9$, normalized to the non-interacting value $w_\text{NI}$. Inset: atomic probe spectra at $\kappa_F a \simeq 0.9$ recorded at 0\,ms (orange squares) and 10\,ms (blue circles) after the quench, compared to that of a non-interacting gas (green crosses). 
The amplitude of each spectrum is normalized to unity and artificially shifted to zero detuning.   
\textbf{(c)} Spectral width $w_{LT}$ at long times normalized to the $t=0$ value $w_{+0}$, as a function of $\kappa_F a$. 
Vertical error bars in all panels combine the standard error of Gaussian fits to the pump and probe spectra with the standard deviation resulting from averaging several data points.
}
\label{Fig4}
\end{center}
\vspace*{-0pt}
\end{figure*}

Let us now discuss the system evolution at longer times $t \gg \tau_F$. While ferromagnetic correlations in the upper branch could foster domain formation over macroscopic length scales at long evolution times \cite{Pekker2011, Sodemann2012, He2016, Zintchenko2016}, our observations rule out this possibility, consistently with Ref.~\cite{Sanner2012}. For $\kappa_F a \gtrsim 1$, despite $\Gamma_{\text{pair}} < \Gamma_{\Delta}$, pairing processes strongly deplete the upper branch population [see Fig.~\ref{Fig2}(c)], yielding $\overline{n}(t)/\overline{n}_{0} < 0.5$ already at $t\sim 1.5$\,ms\,$\sim 50\,\tau_F$ and a corresponding increase of the molecule density. Although pairing slows down at longer times $t > 100\,\tau_F$, it likely prevents the growth of anti-correlations between repulsive fermions over distances beyond few interparticle spacings. The absence of macroscopic spin de-mixing is supported by real-time measurements of the {\it in-situ} density distributions, that reveal only a maximum two-fold increase of local spin-density fluctuations \cite{Recati2010, Sanner2012}, attained after 1\,ms for $\kappa_F a \approx 1$ \cite{SI}. While this moderate enhancement is compatible with the quick formation of microscopic domains, it could also arise from a two-fold local temperature increase within a paramagnetic state \cite{SI}.  

Figure~\ref{Fig4}(a) shows the surviving fermion fraction $\overline{n}_{LT}/\overline{n}_{0}$ (red triangles) together with the relative interaction shift $\Delta_{+LT} /\Delta_{+0}$ (blue circles), recorded within a long-time (LT) interval $10$\,ms\, $\leq\,t\,\leq30$\,ms, as a function of $\kappa_F a$. For $\kappa_F a \geq 1$, we find $\overline{n}_{LT}/\overline{n}_{0} \approx 0.2$, a value essentially steady over more than 100\,ms for $\kappa_F a \simeq 2$.
The combined trends of $\overline{n}(t)$ and spin fluctuations would suggest that chemical equilibrium is reached after some transient time within a hotter, incoherent atom-molecule mixture \cite{Zhang2011, Sanner2012}. Yet, this interpretation conflicts with other observations. A strong suppression of the interaction shift $\Delta_{+LT}$ is found to persist at long times, with $\Delta_{+LT} /\Delta_{+0} < 0.1$ for $\kappa_F a \geq 1$ [see Fig.~\ref{Fig4}(a)].
Neglecting atom-pair interactions, such small $\Delta_{+LT}$ values are inconsistent with those measured in a repulsive Fermi liquid at lower density, even when accounting for a two-fold temperature increase \cite{SI}, implying that the surviving atoms remain indeed anti-correlated. On the other hand, atom-pair interactions, which cannot be reasonably ignored for $\overline{n}_{LT}/\overline{n}_{0} \lesssim 0.2$, could significantly reduce $\Delta_+$ \cite{SI, Jag2014} in high-temperature samples, owing to the compensation between atom-pair $s$-wave repulsion \cite{Petrov2004} and $p$-wave attraction \cite{Levinsen2011} at large collision energies.
However, the scenario of a hot, uncorrelated atom-molecule mixture is irreconcilable with the observed evolution of the atomic peak width $w_{+}(t)$ [see e.g.~Fig.~\ref{Fig4}(b) for $\kappa_F a \simeq 0.9$], which directly reflects the scattering rate of the surviving fermions with the surrounding particles \cite{Jag2014, SI}. This would be substantially enhanced in a hot incoherent sample since, as for the atom-atom case, the atom-dimer collision rate greatly \emph{increases} with increasing collision energy \cite{Levinsen2011, SI}, and would yield $w_{+}(t)$ much larger than the initial $w_{+0}$. 
On the contrary, as directly visible in the probe spectra [see inset of Fig.~\ref{Fig4}(b)], $w_{+}(t)$ rapidly \emph{decreases} below $w_{+0}$, reaching within experimental uncertainty the Fourier-limited width $w_\text{NI}$ of the probe pulse, calibrated on a non-interacting sample.
Fig.~\ref{Fig4}(c) summarizes the long-time behavior of the spectral width $w_{+LT}$, normalized to the initial value $w_{+0}$ obtained for the pump pulse, as a function of $\kappa_F a$.
For $\kappa_F a \geq 1$, $w_{+LT}$ is significantly smaller than $w_{+0}$, matching the Fourier width of our probe pulse within 10$\%$ uncertainty. %
These observations, together with Rabi oscillation measurements \cite{SI}, show that the surviving fermions behave as nearly non-interacting particles, although they remain macroscopically overlapped with the paired component at all times. 
Therefore, the picture of an uncorrelated atom-pair mixture appears too simplistic and qualitatively inconsistent with our data.
Rather, our findings suggest that anti-correlated fermions and pairs coexist within a metastable quantum emulsion \cite{Roscilde2007} or glassy state \cite{Schmalian2000, Dagotto2003}, where the (two) atomic and molecular many-body wavefunctions feature a reduced overlap at short distances, possibly as small as a single interparticle spacing \cite{Zhai2009}.

In conclusion, we studied the evolution of a Fermi gas quenched to strong repulsion. %
We observed the rapid emergence of short-range anti-correlations between itinerant repulsive fermions, dominating the initial dynamics notwithstanding the concurrent emergence of pairing correlations \cite{Pekker2011}. These two competing mechanisms appear equally crucial throughout the entire many-body dynamics.
Our results clarify previous observations, which were ascribed to a ferromagnetic state of atoms \textit{only} \cite{Jo2009}, or to the lack of upper-branch correlations within a rapidly formed atom-molecule incoherent mixture \cite{Sanner2012}.
The observed persistence of strong correlations between repulsive fermions over long times is also consistent with previous studies of spin dynamics at an artificially-created domain wall \cite{Valtolina2017}.  
In the future, it will be interesting to explore the transport properties of such an exotic atom-molecule mixture, and its robustness in weak optical lattices \cite{Pilati2014, Zintchenko2016} or lower dimensions \cite{Conduit2010, Cui2014, Luciuk2017}. Our protocols could also provide exciting possibilities to dynamically access elusive regimes of fermionic superfluidity \cite{Sodemann2012, Bennemann2014}.
%
%
%
%

\smallskip
We acknowledge insightful discussions with Tilman Enss, Tin-Lun Ho, Randall Hulet, Pietro Massignan, Dmitry Petrov, Alessio Recati, Tommaso Roscilde, Christian Sanner and Joseph Thywissen. Special thanks to the LENS Quantum Gases group. This work was supported under European Research Council grants no.~$307032$ QuFerm2D, and no.~$637738$ PoLiChroM, Fondazione Cassa di Risparmio di Firenze project~$2016.0770$ QuSim2D, and by the European Union's Horizon 2020 research and innovation programme under the Marie Sk\l{}odowska-Curie grant agreement no. 705269.

\onecolumngrid
\begin{center}


\newpage\textbf{
\large{Supplemental Material}\\[4mm]
\Large Time-resolved observation of competing attractive and repulsive\\[2mm] short-range correlations in strongly interacting Fermi gases}\\
\vspace{4mm}
{A.~Amico,$^{1,2,*}$
F.~Scazza,$^{1,2,*}$ 
G.~Valtolina,$^{1,2,\dagger}$
P. E. S.~Tavares,$^{1,2,\ddagger}$\\
W.~Ketterle,$^{3}$
M.~Inguscio,$^{1,2}$
G.~Roati$^{1,2}$
and M.~Zaccanti,$^{1,2,\mathsection}$}\\
\vspace{2mm}
{\em \small
$^1$\mbox{LENS and Dipartimento di Fisica e Astronomia, Universit\`{a} di Firenze, 50019 Sesto Fiorentino, Italy}\\
$^2$Istituto Nazionale di Ottica del Consiglio Nazionale delle Ricerche (INO-CNR), 50019 Sesto Fiorentino, Italy\\
$^3$Department of Physics, MIT-Harvard Center for Ultracold Atoms,\\and Research Laboratory of Electronics, MIT, Cambridge, Massachusetts 02139, USA\\}
\vspace{0mm}
{\small$^\mathsection$ E-mail: zaccanti@lens.unifi.it}\\[1mm]
{\small$^*$ These authors contributed equally to this work.}\\
{\small$^\dagger$ Present address: JILA, University of Colorado, Boulder, CO 80309, USA}\\
{\small$^\ddagger$ Present address: Instituto de F\'isica de S\~{a}o Carlos, Universidade de S\~{a}o Paulo, Caixa Postal 369, 13560-970 S\~{a}o Carlos, SP, Brazil}
\end{center}
\setcounter{equation}{0}
\setcounter{figure}{0}
\setcounter{table}{0}
\setcounter{section}{0}
\setcounter{page}{1}
\makeatletter
\renewcommand{\theequation}{S.\arabic{equation}}
\renewcommand{\thefigure}{S\arabic{figure}}
\renewcommand{\thetable}{S\arabic{table}}
\renewcommand{\thesection}{S.\arabic{section}}

\vspace*{15mm}
\twocolumngrid

\section{Experimental procedures}
\subsection{Preparation of the weakly interacting Fermi gas}\label{S1A}
By following procedures described in Refs.~\citenum{Scazza2017, Burchianti2014}, we produce a degenerate $^6$Li Fermi mixture of $N_\text{tot} \approx 10^5$ atoms per spin state, held in a crossed optical dipole trap at a temperature $T/T_F=0.12(2)$. The degenerate gas consists of a balanced mixture of the lowest and third-to-lowest Zeeman states. These are characterized at low magnetic field by quantum numbers \ket{F=1/2,\,m_F=+1/2} and \ket{F=3/2,\,m_F=-3/2}, and they are denoted in the main text and in the following as 1 and 3, respectively. The state labeled as 2 corresponds instead to the hyperfine state \ket{F=1/2,\,m_F=-1/2}.
Efficient evaporation of the 1-3 mixture is achieved by setting the bias magnetic field at 300\,G, where the interspecies scattering length is relatively large, $a_{13}\simeq -900\,a_0$, though non resonant \cite{Burchianti2014}. The final harmonic optical trap is cigar-shaped, with axial and radial frequencies $\omega_\text{ax} \simeq 2\pi \times 20$\,Hz and $\omega_\bot \simeq 2\pi \times 240$\,Hz, respectively. 
To create a population-balanced, weakly interacting $1-2$ mixture, we adiabatically ramp the Feshbach field to 585\,G, where $a_{13} \simeq a_{12} \simeq +300\,a_0$. There, a radio-frequency (RF) pulse of about $150\,\mu$s with optimal frequency and power converts state-3 into state-2 atoms with efficiencies exceeding 98\%. Immediately after, a 3\,$\mu$s-long optical blast selectively removes the leftover state-3 atoms without causing appreciable heating of the remaining $1-2$ sample.

\begin{figure}[h!]
\vspace{10pt}
\includegraphics[width= 8.5 cm]{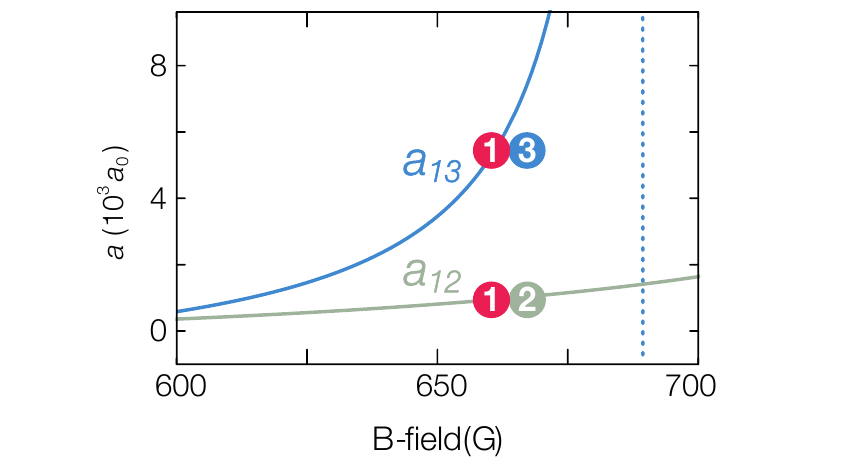}
\caption{Scattering lengths $a_{12}$ and $a_{13}$ between the 1-2 (grey) and 1-3 atomic states (blue), respectively, as a function of the B-field strength. The two off-centered Feshbach resonances at 832\,G and 690\,G allow for tuning $a_{13}$ to large values, while weakly affecting the value of $a_{12}$.}
\label{fig_Feshbach}
\vspace*{-20pt}
\end{figure}

At this point, we increase the bias field to values between 640 and 680\,G, a range where the two off-centered $1-2$ and $1-3$ Feshbach resonances allow for resonantly tuning the scattering length $a_{13}$ (denoted as $a$ in the main text) while only weakly affecting the value of $a_{12}$ \cite{Zurn2013} (see Fig.~\ref{fig_Feshbach}). The final value of the magnetic field is accurately calibrated by driving the $2 \rightarrow 3$ transition in a spin-polarized gas containing only state-2 atoms with a 1\,ms-long RF pulse, and determining its frequency $\nu_0$ with an uncertainty below 30\,Hz. The values of $a_{13}$ at each B-field value are taken from Ref.~\citenum{Zurn2013}, and are used to evaluate the interaction strength for all measurements presented in this work.

\subsection{Effective Fermi energy $\varepsilon_F$ and wavevector $\kappa_F$}
In order to reduce the effect of the density inhomogeneity of trapped samples on the spectroscopy signal, %
we record the transferred atomic fraction by monitoring the column density only within a central rectangular region with size of about 30\,$\mu$m $\times$ 70\,$\mu$m along the transverse and axial directions of the cloud, respectively.
Within this integration region, the system at $t=0$ after the quench is characterized by an effective Fermi energy $\varepsilon_F \equiv \varepsilon_F(t=0)$ and an effective Fermi wavevector $\kappa_F \equiv \kappa_F(t=0)$. $\varepsilon_F$ and $\kappa_F$ represent the mean value of the local $k_F(\br,T/T_F) = (6 \pi^2 n(\br,T/T_F))^{1/3}$ and $E_F(\br,T/T_F) = \hbar^2/(2M)\,k_F^2(\br,T/T_F)$, respectively, averaged over the integration region. These quantities provide the relevant length and energy scales describing our system at the start of the many-body dynamics, and they are used to present our experimental findings in the main text. 

Owing to the finite temperature and the residual trap inhomogeneity, the density of the sample in the integration region differs from the peak density of a zero-temperature Fermi gas, and therefore $\varepsilon_F$ and $\kappa_F$ significantly differ from the zero-temperature central Fermi energy $E_F=k_B T_F= \hbar^2 k_F^2/(2M)=\hbar \, (6N_\text{tot} \,\omega_\text{ax} \,\omega_\bot^2)^{1/3}$ and Fermi wavevector $k_F$. 
We estimate $\varepsilon_F$ (and $\kappa_F$) by numerically evaluating the density-weighted average Fermi energy (wavevector) in the integration region in the local density approximation (LDA) (see also Ref.~\citenum{Scazza2017}). Neglecting the small effect of weak \mbox{1--2} interactions, we approximate the finite-temperature density distribution $n(\br)$ of the initial system with that of a trapped ideal Fermi gas, inserting the measured $T/T_F$, trapping frequencies and total atom number $N_\text{tot}$:
\begin{align}
n(\br) &= - \left(\frac{M k_B T}{2\pi \hbar^2}\right)^{3/2} \text{Li}_{3/2}\left(-\exp\left\{\beta \left(\mu - U(\br)\right)\right\}\right).
\label{localEF}
\end{align}
Here, $\mu(T/T_F, \,N_\text{tot})$ denotes the central chemical potential, $U(\br)$ denotes the harmonic trapping potential, $\beta = 1/(k_B T)$, and Li${}_{s}$ stands for the polylogarithmic function of order $s$.
$\varepsilon_F$ is then obtained by averaging $E_F(\br,T/T_F)$ over the density profile within the integration region, denoted as $V$:
\begin{align}
\varepsilon_F &= \frac{\hbar^2}{2M\,N} \int_V \text{d}\br \left ( 6\pi^2 n(\br) \right )^{2/3} n(\br)\:,
\end{align}
where $N =  \int_V \text{d}\br \,n(\br)$ is the number of atoms in the integration region. Setting $N_\text{tot} = 10^5$ and $T/T_F = 0.12$, we obtain $\varepsilon_F \simeq 0.7\,E_F$ and $\kappa_F \simeq 0.82\,k_F$. Through numerical integration, we also estimate the standard deviation of the local Fermi energy in the chosen region, $\Delta \varepsilon_F \simeq 0.2\,\varepsilon_F$, and the the standard deviation of the local Fermi wavevector, $\Delta \kappa_F \simeq 0.15\,\kappa_F$.  

\vspace*{-7pt}
\subsection{Pump-probe radio-frequency spectroscopy}\label{S1C}
After preparing the system at the target Feshbach field as described in Section \ref{S1A}, we first employ spin-injection (also called \emph{reverse}) radio-frequency (RF) spectroscopy \cite{Gupta2003, Regal2003a, Kohstall2012, Scazza2017} to characterize the spectral response of the balanced, strongly interacting 1--3 mixture. We apply RF pulses with a frequency close to the frequency $\nu_0$ of the $2 \rightarrow 3$ transition (separately calibrated), driving atoms from state 2 to 3 in the presence of the medium of state-1 atoms. 
We employ rectangular RF pulses, whose duration and power are typically adjusted so as to produce a 0.8$\pi$-pulse for a spin-polarized gas. State-2 atoms are thus transferred from the weakly interacting regime to a resonantly interacting one, allowing us to probe the whole many-body spectrum that includes both the ground state and higher-lying excitations such as repulsive quasiparticles (see Fig.~1(b) in the main text). 
Our spectroscopic signal is defined as the ratio between the transferred atoms $N_3$ and the total atomic population $N_2 + N_3$ contained within the integration region $V$ introduced above, monitored as a function of the RF detuning $\delta \nu$ with respect to the bare transition frequency $\nu_0$. The populations $N_2$ and $N_3$ are simultaneously measured for each experimental run by shining two consecutive $5\,\mu$s-long absorption imaging pulses separated by about 500\,$\mu$s from one another and individually resonant with state-2 and 3 atoms at the final B-field. In this way, for each value of the B-field, we determine the detuning $\Delta_{+0}$ associated to the transition from the weakly repulsive state 2 to the strongly repulsive state 3, that corresponds to the positive interaction shift between weakly and strongly repulsive quasiparticle states (see Fig.~2 in the main text and Ref.~\citenum{Scazza2017}). The measured $\Delta_{+0}$ do not depend on the details of the RF pulse, such as power and duration, up to the longest employed pulse of 2\,ms, provided that the pulse duration is kept below that of a $\pi$-pulse for a spin-polarized gas at the same RF power.

The \textit{pump} pulse at a detuning $\Delta_{+0}$ is used to rapidly transfer state-2 atoms to state 3, thereby quenching the repulsive interaction strength. The duration of this pulse is adjusted to maximize the transfer efficiency at the highest power available in our RF circuit. The pump pulse has a duration of approximately 150\,$\mu$s, %
and allows for a transfer efficiency exceeding 95\% at moderate couplings $\kappa_F a < 1$. To quench the system to stronger couplings, the duration of the pump pulse duration has to be adjusted to up to 250\,$\mu$s for $\kappa_Fa \gtrsim 2$. This is due to an increased dressing of the repulsive quasiparticles that reduces the Franck-Condon overlap between the initial and the final state, i.e.~the quasiparticle residue, thereby reducing the associated Rabi coupling at a given RF power \cite{Scazza2017, Kohstall2012}. In addition, the increased collisional rate between state-3 fermions and the state-1 Fermi sea causes an appreciable damping of the Rabi oscillation: this results in an overall reduced attainable transfer on the $2 \rightarrow 3$ transition, whose efficiency remains however above 70\% even for the strongest interactions explored in this work. 
In order to purify the final sample, we remove the residual state-2 population by means of a spin-selective 3\,$\mu$s resonant optical blast, applied directly at the end of the pump pulse. The intensity of the optical blast is carefully adjusted to yield negligible heating and losses in the interacting 1-3 mixture, while ensuring complete removal of the undesired state-2 population. 
We remark that an imperfect transfer during the pump pulse may cause an intrinsic heating of the state-3 Fermi gas, even in the absence of inelastic pairing processes. 
This stems from the fact that the coherence between state-2 and state-3 fermions established by the RF pump pulse is zeroed upon resonantly removing the state-2 atomic population from the trap. The RF transfer procedure results in a state-3 atom number that is smaller than the initial state-2 one, leading to a reduced state-3 Fermi energy and thus to an increased $T/T_F$. 
For our initial temperature of $T/T_F=0.12$, such a heating mechanism is expected to cause an appreciable though not substantial decrease of the gas degree of degeneracy, given our typical transfer efficiencies. 
For instance, for a transferred fraction of 85\% (70\%), typical for $\kappa_F a \sim 1.5$ ($\kappa_F a \sim 2.2$) we expect the state-3 fermions to reach $T_{\text{fin}}/T_{F,3} \simeq 0.17$ ($T_{\text{fin}}/T_{F,3} \simeq 0.20$). This estimate is obtained assuming that the populations of the two internal atomic states thermalize in the harmonic trap to the same final temperature $T_{\text{fin}}$, and recalling that the final Fermi temperature of the state-3 component is given by $T_{F,3}=\alpha^{1/3}\,T_F$, $\alpha$ being the transferred fraction.

The \textit{probe} pulse is used to monitor spectroscopically the evolution of various quantities after the repulsion quench in the $1-3$ mixture: the interaction shift $\Delta_+(t)$, the relative atomic population $\overline{n}(t)/\overline{n}_{0}$ and the relative molecular population $\overline{n}_{mol}(t)/\overline{n}_0 \equiv 1 - \overline{n}(t)/\overline{n}_{0}$. This spectroscopy pulse has a typical length of 250 to 500\,$\mu$s, and its power level is set to that of a free-atom $\pi$-pulse with the same duration. 
The relative atomic upper-branch population $\overline{n}(t)/\overline{n}_{0}$ is linked to the amplitude $A(t)$ of the atomic spectrum as $\overline{n}(t)/\overline{n}_{0} = A(t)/A(0) = A(t)/\alpha$. Here, $\alpha \leq 1$ denotes the atom transfer efficiency of the $3 \rightarrow 2$ probe pulse at $\Delta_+(t)$. 
It is assumed to coincide with the efficiency of an identical $2 \rightarrow 3 $ pump pulse, which allows to precisely calibrate the atom transfer, free from a molecular fraction contaminating the spectroscopy signal.  

In principle, the time-dependent paired fraction $\overline{n}_{mol}(t)/\overline{n}_0$ could alternatively be extracted by monitoring the rising amplitude of the molecular spectral response at $\delta \nu < 0$. However, the pair dissociation efficiency is strongly interaction-dependent, due to a decreasing Franck-Condon overlap with decreasing $\kappa_F a$ . Hence, a quantitative extraction of the pairing rate from the broad molecular signal becomes increasingly challenging for smaller interaction strengths $\kappa_F a < 1$. 

For the measurement of $\Gamma_{\Delta}$ and $\Gamma_\text{pair}$ at strong coupling, probe pulses as short as 50\,$\mu$s have been employed.  
In this latter case, the pulse power is adjusted to the maximum one attainable with our RF circuit, resulting in about a $\pi/3$-pulse for non interacting atoms.  
Although this unavoidably reduces our spectral resolution and diminishes the signal-to-noise ratio of our measurements, it enables to considerably increase the time resolution of our spectroscopy, and to trace the very fast dynamics of the strongly repulsive gas over timescales as short as $\sim 2\,\tau_F$. 
The RF spectra, obtained by recording the transferred atomic fraction on the $3 \rightarrow 2$ transition upon applying the probe pulse at variable detuning, contain spectral features both at positive and negative detunings, associated with (attractive) molecular and (repulsive) atomic states, respectively. At coupling strengths $\kappa_F a \leq 1.6$ the two spectral features are well separated, and a Gaussian fit to the atomic resonance at $\delta\nu >0$ is sufficient to extract its center $\Delta_+(t)$ and amplitude $A(t)$.
Conversely, at strong couplings $\kappa_F a > 1.6$ the two features partially overlap and the resulting spectra are well fitted by the sum of a Gumbel and a Gaussian function, respectively accounting for the molecular and atomic contributions (see Fig.~1(c) in the main text and Fig.~\ref{fig_atom_mol_profiles}). In this latter case, the Gumbel is a phenomenological model spectrum and serves only to subtract the molecular background from the unpaired atomic signal.
From the center and amplitude obtained from the Gaussian fit of the atomic spectra for evolution times $t\leq 200\,\mu$s, $\Gamma_{\Delta}$ and $\Gamma_\text{pair}$ are determined from a linear fit to the normalized data: $\Delta_+(t)/\Delta_{+0}=1-\Gamma_{\Delta}$t, and $\overline{n}(t)/\overline{n}_{0}=1-\Gamma_\text{pair}$\,t, respectively.
Notwithstanding the moderate spectral resolution associated to a width of at least 0.2\,$\varepsilon_F(0)$ resulting from Fourier broadening of the probe pulse, $\Delta_+(t)$ is determined with a typical standard fitting error lower than 50\,Hz (see Fig.~1(c) in the main text).

\section{Three-body recombination rate for balanced Fermi mixtures} 
As discussed in the main text, one major finding of our work is the fact that pairing processes appear to develop through three-body recombination processes, rather than via the (much faster) Cooper-channel pairing considered in Ref.~\cite{Pekker2011}.
Three-body processes are responsible for the recombination of two fermions into a bound-state dimer within a scattering event involving a third atom, the latter ensuring both energy and momentum conservation during the inelastic collision. 
As long as the mean collision energy of the particles remains lower than the binding energy of the dimer, and the system is far from the unitarity limit $\kappa_F a \gg 1$, the rate for three-body processes can be quantitatively described by the theoretical approach developed by Petrov in Ref.~\cite{Petrov2003}. 

For the equal mass, broad resonance case relevant in this work, the number of recombination events of the kind \mbox{$\uparrow+\uparrow+\downarrow \,\,\,\,\longrightarrow\,\,$}\mbox{$(\uparrow \downarrow)$}\,$+ \uparrow$ per unit of time and volume reads~\cite{Petrov2003}:
\begin{align}
\alpha_{\uparrow \uparrow  \downarrow } n_{\uparrow}^2=148 \frac{\hbar a^4}{m} \frac{\overline{\varepsilon}}{\varepsilon_b}n_{\uparrow}^2
\label{3B_1}
\end{align}
Here, $n_{\uparrow}$ denotes the density of $\uparrow$ fermions, $\overline{\varepsilon}$ their mean kinetic energy, and $\varepsilon_b$ stems for the energy gained by binding two fermions into a dimer. In the presence of two balanced Fermi seas of $\uparrow$ and $\downarrow$ particles with Fermi energy $\varepsilon_F$, this latter reads $\varepsilon_b= \hbar^2/(m a^2)+2 \varepsilon_F$.
In the population-balanced case, $n_{\uparrow 0}=n_{\downarrow 0}\equiv n_0$, the alternative recombination channel  $\downarrow + \downarrow + \uparrow \,\,\,\, \longrightarrow  \,\,\,\, (\uparrow \downarrow) + \downarrow$ is equally important, and the number of events of this kind is given by Eq. \eqref{3B_1} with $\uparrow$ and $\downarrow$ interchanged, being $\alpha_{\uparrow \uparrow  \downarrow }=\alpha_{\downarrow \downarrow  \uparrow }=\alpha$.

For most of the interaction strengths explored in our work, $\kappa_F a \geq 0.5$, both the dimer and the atom resulting from one three-body collision event remain confined in the optical dipole trap. Hence, at each spatial position of the sample the fermion density $n(t)$ follows the rate equation:
\begin{align}
\dot{n}(t)=-2 \times 148 \frac{\hbar a^4}{m} \frac{\overline{\varepsilon}}{\varepsilon_b}n^3(t).
\end{align} 
At short times, $n(t) \simeq n_0$, this can be approximated as
\begin{align}
\dot{n}(t) \simeq -2 \times 148 \frac{\hbar a^4}{m} \frac{\overline{\varepsilon}}{\varepsilon_b}n_{0}^2\,n(t)
\label{3B_3}
\end{align}
yielding the initial three-body decay rate  
$\Gamma_3 \simeq 2 \times 148 \, \frac{\hbar a^4}{m} \, \frac{\overline{\varepsilon}}{\varepsilon_b} \, n_{0}^2$.
Expressing the fermion density in terms of the (local) Fermi wavevector, and normalizing $\Gamma_3$ to the Fermi energy $\varepsilon_F$, we obtain the following expression for the rate:
\begin{align}
\frac{\hbar \Gamma_3}{\varepsilon_F}= 2 \times  0.806  \times \frac{148}{(6 \pi^2)^2} \frac{(\kappa_F a)^6}{1+(\kappa_F a)^2},
\label{3B_4}
\end{align}
where the numerical factor $0.806$ results from averaging the kinetic energy and the squared fermion density over the integration region.  
\refeq{3B_4} is shown as a line in the inset of Fig. 3 of the main text, and it is dashed in the strongly interacting regime, where unitarity limited interactions may modify the result of the isolated three-body problem.

%
\section{Impact theory of\\pressure-induced effects on spectral lines}
A major finding of our study, highlighted in Fig. 4 of the body of the paper, is the suppression of interaction shifts at long evolution times and strong couplings, $\Delta_{+LT}/\Delta_{+0}\leq 0.1$, combined with spectral widths of the atomic peak $w_{+LT}$ that approach the Fourier width of the probe pulse. 
In the following we provide additional information to better understand our interpretation of long-time data.

\begin{figure}[t!]
\vspace*{-10pt}
\includegraphics[width= 9cm]{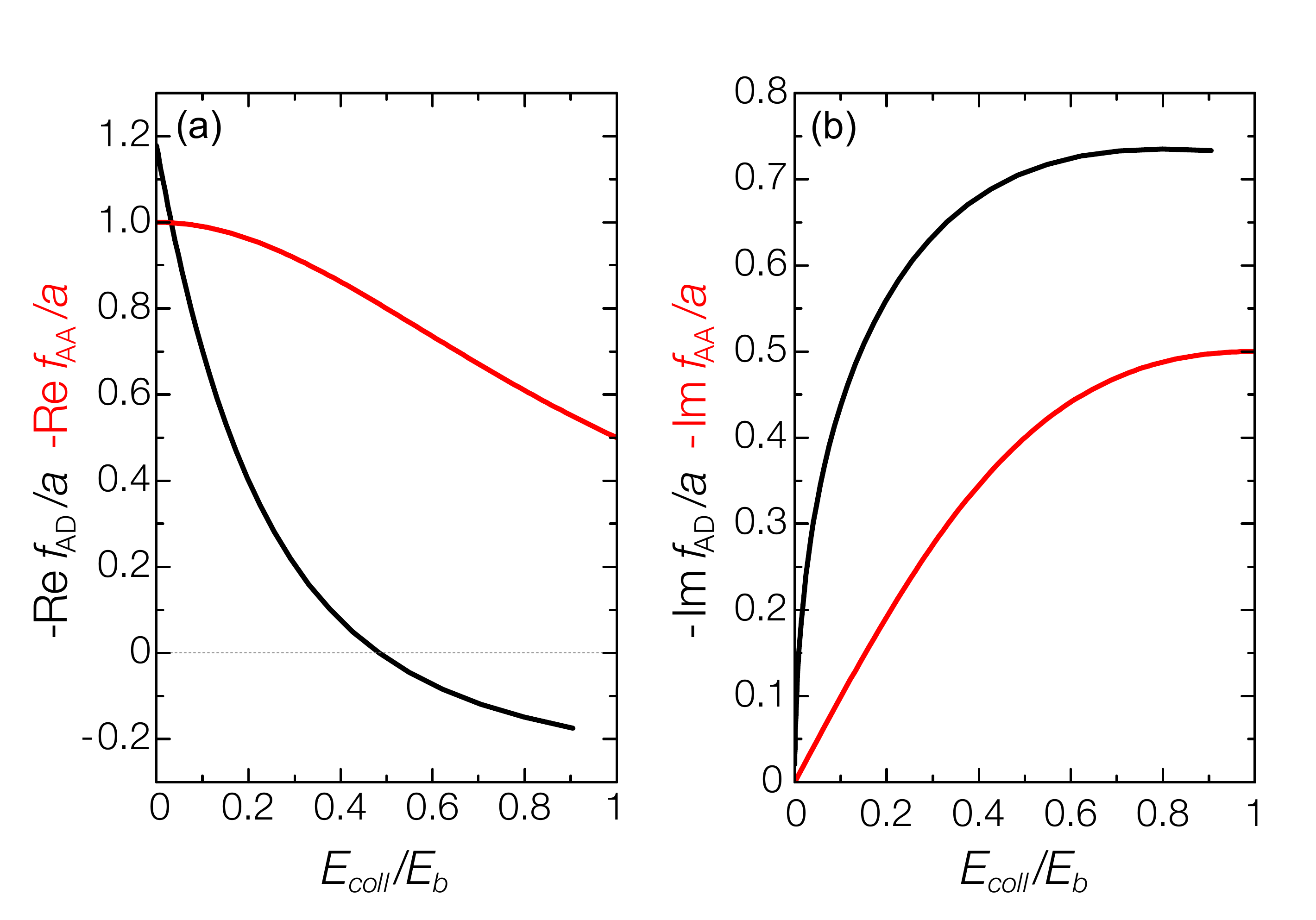}
\caption{(\textbf{a}) Real part of the atom-dimer (black line) and atom-atom (red line) forward scattering amplitudes, normalized to the atom-atom scattering length $a$, as a function of the collision energy $E_{coll}$ normalized to the dimer binding energy in the vacuum $E_b$. The atom-dimer curve only includes the dominant $s$- and $p$-wave contributions to the total scattering amplitude. The weak repulsive $d$-wave contribution, here not included, although leading to slight quantitative corrections, does not change the qualitative trend of $-{\rm Re} f_{\rm AD}$. (\textbf{b}) Imaginary part of the atom-dimer (black line) and atom-atom (red line) forward scattering amplitudes, normalized to the atom-atom scattering length $a$, as a function of the collision energy $E_{coll}$ normalized to the dimer binding energy in the vacuum $E_b$. Also in this case, the atom-dimer curve neglects $l$-wave contributions with $l\geq 2$ to the total scattering amplitude, which would cause a slight further increase of ${\rm {\rm Im}} f_{\rm AD}$, relative to ${\rm {\rm Im}} f_{\rm AA}$. 
}\label{fAD}
\vspace*{-10pt}
\end{figure}

As discussed in the main text, at long evolution time the system for $\kappa_F a\geq 1$ is comprised of an atom-pair mixture, in which the surviving fermions represent the minority component, $\overline{n}(t)/\overline{n}_0 \leq 0.2$. Parallel to this, temperature estimates based on the study of spin density fluctuations and on the analysis of the in-situ density profiles of the atomic and molecular components consistently yield a long-time, stationary value two-to-four times higher than the initial temperature (see Section \ref{insitu}).
As such, one is tempted to regard the resulting system as an uncorrelated, thermal mixture of (minority) fermions and (majority) pairs. Such a regime was already explored in the case of mass-imbalanced K-Li Fermi mixtures at narrow resonances \cite{Jag2014}, and the resulting spectra were successfully analyzed and interpreted on the basis of the impact theory of pressure-induced effects on spectral lines. 
In this framework, the collisions of the fermions with the surrounding bath are considered to be effectively instantaneous. The resulting shift and full width at half maximum (FWHM) of the atomic spectra can be linked to the real and imaginary parts of the forward atom-dimer scattering amplitude \mbox{$f_{\rm AD}$,~respectively,~as:}
\begin{align}
	h\Delta_+=\frac{2\pi \hbar^2}{\mu} {\rm Re} \left\langle f_{\rm AD}\right\rangle n_{\rm D},\\
	h w_+=2 \frac{2\pi \hbar^2}{\mu} {\rm Im} \left\langle f_{\rm AD}\right\rangle n_{\rm D},
\end{align}
where $\mu=\frac{2}{3}m$ is the atom-dimer reduced mass, $n_{\rm D}$ the molecule density, and $\left\langle f_{\rm AD} \right\rangle$ denotes the thermodynamic average of the $f_{\rm AD}$ over all atom-dimer collision energies $E_{coll}~=~\hbar^2 k_{coll}^2/2\mu$.

In contrast with the case of atom-atom collisions in the cold regime, fully characterized by the $s$-wave scattering amplitude, atom-molecule scattering is strongly affected by higher-order partial waves contributions, in particular those associated with the $p$-wave channel $l=1$ \cite{Levinsen2011}.
Neglecting $l\geq2$ contributions, the resulting forward scattering amplitude for atom-dimer collisions reads \cite{Jag2014}
\begin{eqnarray*}
	f_{\rm AD}(k_{coll})= \frac{\sin(2\delta_0(k_{coll}))}{2 k_{coll}}+3\frac{\sin(2\delta_1(k_{coll}))}{2 k_{coll}}+\\
	+\, i \left[\frac{\sin^2(\delta_0(k_{coll}))}{k_{coll}}+ 3 \frac{\sin^2(\delta_1(k_{coll}))}{k_{coll}}\right],
\end{eqnarray*}   
where $\delta_0$ and $\delta_1$ represent the $s$- and $p$-wave atom-dimer phase shifts, respectively, which have been calculated by Levinsen and Petrov in Ref.~\cite{Levinsen2011} also for the equal mass, broad resonance case.
The resulting trends for the real and imaginary parts of $f_{\rm AD}$, normalized to the atom-atom scattering length $a$, are shown as black lines in Fig.~\ref{fAD}(a) and (b), respectively. 
For convenience, they are plotted as a function of $E_{coll}/E_b$ with $E_b=\hbar^2/(m a^2)$. For comparison, we also show the corresponding trend for atom-atom $s$-wave collisions (red lines), given by the scattering amplitude $f_{\rm AA}$:
\begin{eqnarray*}
	{f_{\rm AA}(k_{coll}) \over a}=-\frac{1}{1+ (k_{coll} a)^2}+i \frac{k_{coll}a}{1+ (k_{coll} a)^2}
\end{eqnarray*}      
From Fig.~\ref{fAD}(a) one can notice how $-{\rm Re} f_{\rm AD}$ starkly differs from the trend exhibited by $-{\rm Re} f_{\rm AA}$ as the collision energy is increased to sizable fractions of $E_b$: $-{\rm Re} f_{\rm AA}$ shows only a slow decrease with increasing $E_{coll}$ while remaining positive, i.e. leading to an atom-atom repulsive interaction. In contrast, $-{\rm Re} f_{\rm AD}$ features a much sharper drop, becoming zero for $E_{coll}/E_b\sim 0.5$, and turning negative for even larger collision energies. This peculiar behavior arises from the $p$-wave atom-dimer attractive interaction term, overcoming the repulsive $s$-wave one for sufficiently large collision energies \cite{Levinsen2011, Jag2014}. 
From Fig.~\ref{fAD}(a) alone, one could conclude that the interaction shifts $\Delta_{+LT}$ recorded at long evolution times may indeed arise with atom-dimer interactions at (averaged) collision energies on the order of 0.5$E_b$. 

On the other hand, Fig.~\ref{fAD}(b) rules out the possibility that the observed long-time dynamics is due to atom-molecule collisions in an incoherent mixture. In contrast with $-{\rm Re} f_{\rm AD}$, ${\rm Im} f_{\rm AD}$ monotonously increases up to the breakup threshold, $E_{coll}/E_b=1$, featuring values almost twice as large as the ones expected for atom-atom collisions at the same energy (and density). This trend results from the combined $s$- and $p$-wave contributions to the total scattering cross section, that are both positive.
Since ${\rm Im} f_{\rm AD}$ is directly linked to the spectral width, atom-dimer collisions are expected to yield a spectral width significantly larger than the initial one $w_{+0}$. 
Indeed, an incoherent molecular bath with density equal to the initial one $n_0$ at a temperature $T_{\rm fin}$ two-to-four times higher than the initial one, would feature an average collision energy $E_{\rm fin}\sim 3/2\,k_B T_{\rm fin}$. This corresponds to about 2 times the initial collision energy relevant in the (atom-atom) scattering, $E_{in}\sim 3/8\,k_B T_{F}$. Irrespective on the specific value of $E_{coll}/E_b$, from Fig.~\ref{fAD}(b) one should then expect, at long time, spectral widths $w_{+LT}$ at least a factor two higher than the initial one. Since the $w_{+LT}$ observed in our experiments is almost that of a non-interacting, spin-polarized sample, we conclude that incoherent atom-dimer interactions cannot be reconciled with our data.

\section{Additional analysis of spectroscopic data}

\subsection{Mass renormalization effects in spin-injection spectroscopic data}
As discussed in the main text, the observation of a coherent atomic peak at positive detuning $\Delta_{+0}$ in the spin-injection spectroscopy demonstrates the existence of a well-defined repulsive branch of the resonant 1-3 mixture.
The measured interaction shifts encode information about the properties of the quasi-particles composing the repulsive Fermi liquid. In the following, we further elaborate on the spin-injection spectroscopic data, connecting the results obtained in the spin-balanced regime with the ones recorded in the impurity limit \cite{Scazza2017}. 
In general, the link between the spin-imbalanced polaron scenario and the spin-balanced Fermi liquid is not straightforward \cite{Pilati2010}. In the balanced case, the properties of the associated quasi-particles, such as their zero-momentum energy and effective mass, may strongly differ from the ones associated with the case of a single impurity embedded in a Fermi gas. Moreover, effective quasi-particles interactions can be relevant in the balanced case, while being negligible in the limit of vanishing impurity concentration. 

In a previous study \cite{Scazza2017}, we indeed observed a strong dependence of the RF shift $\Delta_{+0}$ upon the impurity concentration $x$, with $\Delta_{+0}$ decreasing considerably when increasing $x$ at strong coupling. 
Rather than being connected with polaron-polaron interactions, this peculiar trend was found to arise from the different dispersions featured by the initial weakly-interacting impurities (of bare mass $m$) and the final quasiparticles (with effective mass $m^*>m$). 
As the impurity concentration is increased at a fixed temperature, the mean motional energy per impurity $\overline{\epsilon}$ grows non-linearly, owing to the increased Fermi pressure of the minority gas (see details in Ref.~\cite{Scazza2017} and Supplemental Material). As a consequence, the resulting shifts acquired at given $\kappa_F a$ and $0.05\leq x \leq 0.4$ were found to depend linearly upon the impurity mean energy  $\overline{\epsilon}$ with a negative slope directly reflecting the value of $m/m^*$:
\begin{align}
	\Delta_{+0} = E_+ - (1-\frac{m}{m^*})\overline{\epsilon}
\end{align}
Here, $E_+$ stands for the energy of a zero-momentum repulsive quasi-particle, i.e. a repulsive polaron. 
For the same reason, $\Delta_{+0}$ saturates as a function of $\kappa_F a$ at fixed concentration [see e.g.~blue squares in Fig. 2(a) of the main text].
The trend of $\Delta_{+0}$ measured in this work by spin-injection spectroscopy in balanced mixtures appears also consistent with the effect of quasi-particle mass renormalization, with $\Delta_{+0}$ saturating at strong couplings in a similar manner as in the impurity limit [see circles in Fig.~2(a) of the main text].

\subsection{Incompatibility of the long-time spectral response with simple density reduction of the Fermi liquid}
As discussed in the main text, at strong repulsion and long evolution times our spectroscopy data unveil the persistence of a small atomic fraction featuring negligible interaction shift and collision-induced spectral broadening (see Fig.~4 of the main text). Based on the observed long-time behavior of the widths of the atomic peaks, in the main text we concluded that the overlap between atomic and molecular many-body wavefunctions must be strongly reduced. 
In the following, we provide further analysis of our time-resolved probe spectroscopy data, which appear to exclude the possibility that the observed mid- and long-time behavior of $\Delta_+(t)$ could be ascribed to a repulsive Fermi liquid with $n_1(\textbf{r}) \approx n_3(\textbf{r})$ at higher temperature and/or reduced overall density (neglecting atom-pair interactions).

\begin{figure}[t]
\includegraphics[width= 0.9\columnwidth]{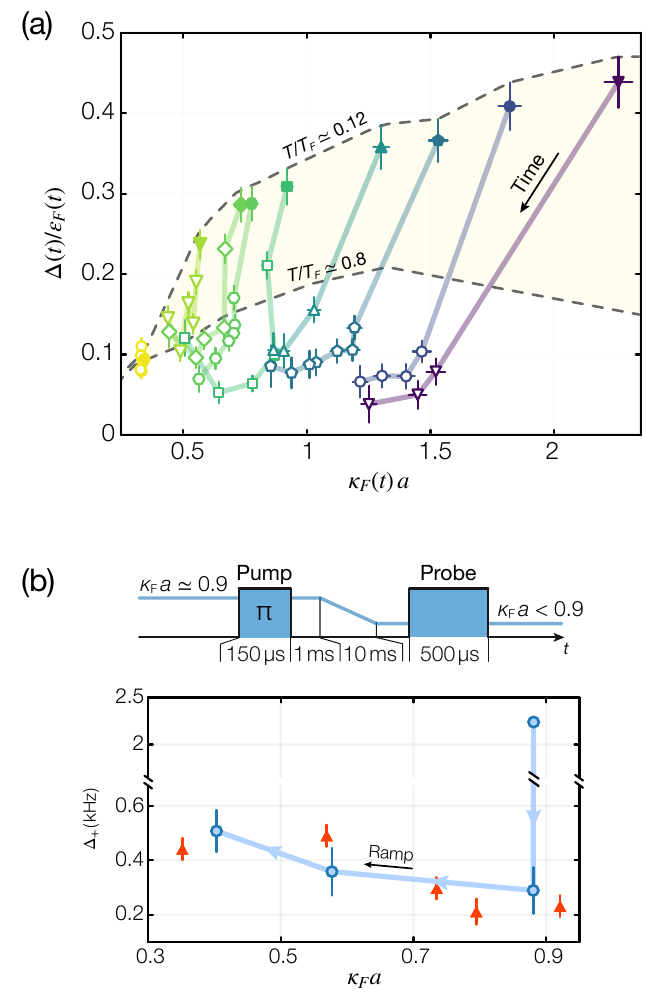}
\caption{({\bf a}) $\Delta_+(t)/\varepsilon_F(t)$ versus $\kappa_F(t)\,a$ for a system initially at $T/T_F=0.12(2)$. The upper (lower) limit of the shaded region is obtained via spin-injection spectroscopy on a paramagnetic sample at $T/T_F=0.12(2)$ ($T/T_F=0.80(5)$). Vertical (horizontal) error bars account for the uncertainty in determining $\Delta_+(t)$ and $\varepsilon_F(t)$ ($\kappa_F(t)$). ({\bf b}) A 10 ms-long magnetic field ramp to lower $(\kappa_Fa)_{\rm f}$ re-establishes an interaction shift (blue circles) compatible with $\Delta_{+LT}$ obtained for a gas quenched at low $\kappa_F a$ (red triangles).}
\label{fig_FM}
\vspace*{-10pt}
\end{figure}

In order to compare our long time observation with this latter possibility, we exploit the knowledge of $\Delta_{+0}$ acquired through the spin injection spectroscopy data at $T/T_F=0.12(2)$ and $T/T_F=0.80(5)$, shown in Fig.~2(a) of the main text.
The results of this characterization at low and high temperature delimit the shaded area in Fig.~\ref{fig_FM}(a) from top and bottom, respectively. 
As done already in the main text, for each temperature, $\Delta_{+0}$ is normalized to the relevant $\varepsilon_F$, and plotted as a function of the associated $\kappa_F a$. 
It is important to stress how a temperature $T/T_F \sim 0.8$ corresponds to an average collision energy greatly exceeding the one estimated for the gas throughout the dynamics after the quench at all interaction strengths.
As such, if the system remained in the non-correlated Fermi liquid phase at long evolution times, the associated interaction shifts should definitely fall within the shaded region of Fig.~\ref{fig_FM}(a).

In order to account for the drop of the unpaired atomic density, we plot the evolution of the interaction shift $\Delta_+(t)$ as function of the \textit{instantaneous} interaction strength $\kappa_F(t) \,a$ and normalize it to the instantaneous Fermi energy $\varepsilon_F(t)$. 
The estimate for the instantaneous Fermi energy and wavevector is based on the experimental knowledge of $\overline{n}(t)/\overline{n}_0$: 
\begin{align*}
\kappa_F(t)&= \kappa_F\cdot (\overline{n}(t)/\overline{n}_0)^{1/3}\\
\varepsilon_F(t)&=\varepsilon_F\cdot (\overline{n}(t)/\overline{n}_0)^{2/3}
\end{align*}
The resulting trends for various initial interaction strengths $\kappa_F a$ are presented in Fig.~\ref{fig_FM}(a).
If the surviving fermions remained in the non-correlated Fermi liquid state, possibly at a temperature higher than the initial one, each data set would  fall within the shaded region. The combination of reduced atom density and heating can therefore explain the dynamics observed at small $\kappa_F a$, but appears irreconcilable with the trend observed at $\kappa_F a \geq 0.8$. There, the drop of $\Delta_+(t)$ is much larger than that expected from the trivial depletion of atom density. 
This leads to the conclusion that, for $\kappa_F a \geq 0.8$, the repulsive Fermi gas relaxes onto a state featuring 
a strong decrease of the density overlap
between unpaired atoms occupying different spin states. %
The non-trivial nature of the atom-dimer mixture reached at long times is further supported by the reversibility of the interaction shift. As displayed in Fig.~\ref{fig_FM}(b), upon ramping down the interaction strength to $(\kappa_Fa)_{\rm f}$ in 10\,ms after 1\,ms of evolution at $\kappa_Fa \simeq 0.9$, $\Delta_+$ increases (blue circles) and reaches within the experimental uncertainty the value $\Delta_{+LT}$ obtained at long evolution time for a system quenched to $(\kappa_Fa)_{\rm f}$ (red triangles).

\subsection{Spatial dependence of $\Delta_+(t)$}
\begin{figure}[t]
\centering
\includegraphics[width= 8.6 cm]{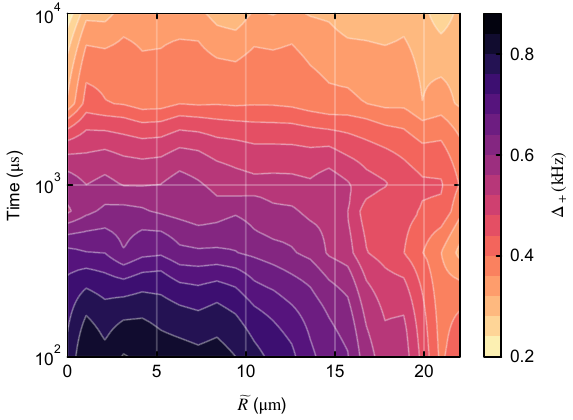}
\caption{\label{slices}
Spectral response of a $1-3$ gas mixture as a function of evolution time and distance $\tilde{R}$ from the trap center for $\kappa_F\,a \sim$ 1.2. The RF shift $\Delta_+$ is initially much stronger in the center than in the external regions of the cloud. However, as the system evolves into the strongly-repulsive state, $\Delta_+(t)$ within the central portion of the cloud quickly drops to values as low as 200\,Hz. In contrast, the outer wings of the sample present a much slower and less pronounced evolution of the interaction shift, consistent with a pure drop of the atom density. Intermediate regions, $10 \leq \tilde{R} \leq 20\,\mu$m, present a trend smoothly crossing-over between these two limiting behaviors.}
\vspace*{0pt}
\end{figure}

In this section we provide further analysis of our spectroscopy data, highlighting the observed $\Delta_+(t)$ dynamics within different spatial regions of the trapped cloud, at increasing distance $\tilde{R}$ from the trap center.
In Fig.~\ref{slices}, we present the evolution of the interaction shift after the repulsion quench at varying radial distance from the trap axis for a two-component 1-3 gas quenched at $\kappa_F a \sim 1.2$.
As one can notice, different regions with different particle densities initially experience an interaction shift which monotonically decreases as one moves out from the denser central region.
As already discussed in the main text, while spin anti-correlations are established in the system, $\Delta_+(t)$ rapidly drops to values as low as 10-20$\%$ of the initial $\Delta_{+0}$ for $\tilde{R}\leq 10\,\mu$m. 

On the other hand, outer regions of the cloud, $\tilde{R} \geq 20\,\mu$m, owing to a significantly lower local initial density and hence lower interaction strength, exhibit only a weak drop of the RF shift $\Delta_{+LT}/\Delta_{+0} \leq 2$, ascribable to a mere density reduction and heating. 
In this outer shell of the gas, ferromagnetic correlations are absent at all evolution times. Intermediate regions, $10 \leq \tilde{R} \leq 20\,\mu$m, smoothly cross-over between these two extremal behaviors, presenting intermediate relative drops occurring at progressively slower rates. 

\subsection{Rabi oscillations in the anti-correlated regime}

\begin{figure}[t!]
\centering
\includegraphics[width= 8.5 cm]{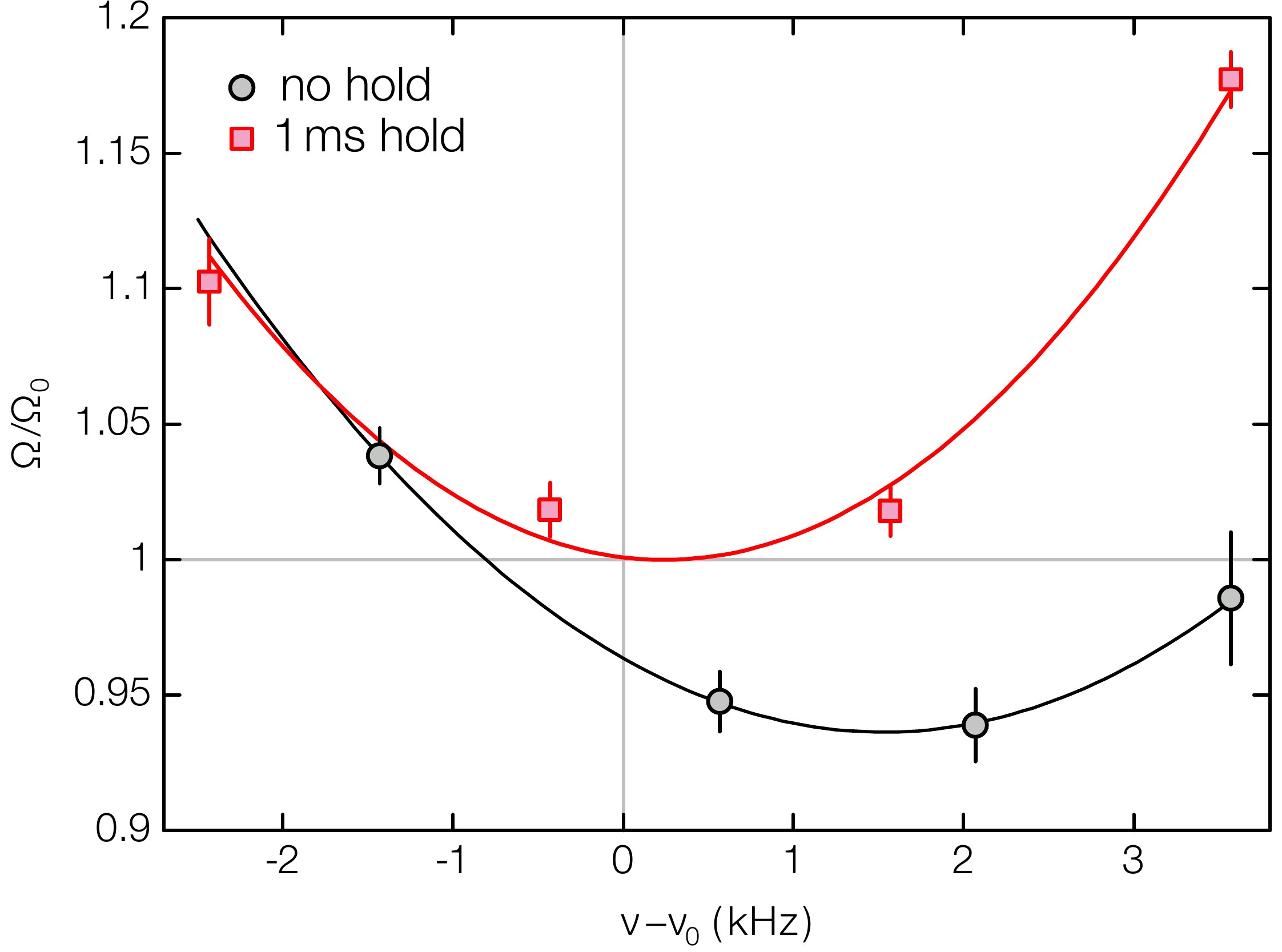}
\caption{\label{figRabi}
Behavior of the Rabi frequency versus detuning from the bare-atom resonance at $\kappa_Fa \sim$ 2, both in the anti-correlated state (red squares), and in the Fermi liquid state (gray circles). 
To reach the anti-correlated regime, we let the system evolve for $t=$1\,ms after the quench, subsequently driving Rabi oscillations on the $2 \rightarrow 3$ transition with a detuning $\Delta_+(t)$ corresponding to the location of the atomic peak. 
The Rabi frequency for the Fermi liquid case is obtained instead by simply performing Rabi oscillations starting from a 1-2 weakly-interacting mixture. 
In both cases the RF power is set to the maximum value allowed by our RF circuit, resulting in free-atom Rabi frequencies of about 5.2\,kHz.
Atoms driven on the $2 \rightarrow 3$ transition in the paramagnetic state feature a considerably strong renormalization of the frequency, $\Omega <\Omega_0$, which reaches its minimum at $\Delta_{+0} \sim 2\,$kHz. In turn, atoms in the anti-correlated state oscillate at the bare atom Rabi frequency, with the minimum frequency found near zero detuning, $\delta \nu = \nu - \nu_0 \simeq 200$\,Hz.
}
\vspace*{-0pt}
\end{figure}

As discussed in the main text, the persistence of ferromagnetic correlations at strong coupling even in the presence of an increasingly large paired component are strongly suggested by the observed strong drop of the widths of the atomic spectra $w_{+LT}$ recorded at long evolution times. This interpretation of the long-time dynamics is also supported by the study of Rabi oscillations, not discussed in the main text and here briefly described. 
For this measurement we set the RF frequency at the previously determined $\Delta_+(t)$ while the RF power is set to the maximum value allowed by our apparatus. 
Similarly to what done for the investigation of repulsive polaron Rabi oscillations in Ref.~\citenum{Scazza2017}, we extract the frequency $\Omega$ and the damping rate $\gamma_R$ by fitting the data with the function $f(t) = A\,e^{-\Gamma_R t} + B\,e^{-\gamma_R t} \cos(\Omega t)$, describing a Rabi oscillation at frequency $\Omega/(2\pi)$ with a damping $\gamma_R$ and an excited-state population decay rate $\Gamma_R$ (with $A, B \simeq 0.5$). 
In the repulsive Fermi liquid regime, the renormalization of the quasiparticle coherence, encoded in a quasiparticle weight smaller than unity, results in $\Omega/\Omega_0 \leq 1$ \cite{Scazza2017, Kohstall2012}. 
Here, $\Omega_0$ denotes the free-atom Rabi frequency, experimentally measured using a spin-polarized state-2 Fermi gas.
In turn, when short-range correlations emerge in our system, resulting in $\Delta_+(t) \sim 0$, the coherent oscillations occur at a Rabi frequency matching within the experimental uncertainty that of a spin-polarized gas, $\Omega/\Omega_0=1$, yet featuring a finite damping. 
This trend of the Rabi frequency is summarized in Fig.~\ref{figRabi} for $\kappa_Fa \sim 2$: here we compare $\Omega/\Omega_0$ extracted by monitoring the $2 \leftrightarrow 3$ Rabi flopping at different detunings $\Delta$ of the RF drive from the non-interacting transition frequency, on a repulsive Fermi liquid (gray circles) or a gas within which anti-correlations have developed (red squares), respectively.
The latter trend is obtained by first quenching the system in the interacting regime and letting it evolve for 1\,ms, prior to start the acquisition of the Rabi oscillation signal.
In both cases, the Rabi frequency quantitatively matches the usual trend $\Omega_{\Delta} \sim \sqrt{\Omega^2+\Delta^2}$ expected by coherently coupling two discrete energy levels. Note that in the Fermi liquid (anti-correlated) phase the minimum value $\Omega/\Omega_0<1$ ($\Omega/\Omega_0=1$) is reached at a positive (zero) detuning from the free atom transition resonance.
This observation further supports the emergence of repulsive fermion anti-correlations and ferromagnetic fluctuations discussed in the main text, which make state-3 fermions behave as non-interacting particles with quasiparticle weight compatible to 1. 
On the other hand, the sizable damping of such coherent oscillations (not shown), suggests that ferromagnetic correlations may melt during the Rabi cycles, allowing for collisional decoherence effects to develop. This also rules out the presence of macroscopic domains extending over several interparticle spacings, for which Rabi decoherence times exceedingly longer than those measured on a repulsive Fermi liquid would be expected.  

\section{Analysis of in-situ density profiles} \label{insitu}
In this Section we provide additional data related to the dynamics of the in-situ density distributions of the atomic and molecular components. As mentioned in the main text, the analysis of the in-situ images enables to provide an estimate for the system temperature at long evolution times, and to rule out any signature of macroscopic phase separation between the paired phase and the surviving fermions in our trapped sample. 
First of all, it is worth remarking that macroscopic phase separation between pairs and fermionic atoms has been widely investigated in the context of trapped, polarized superfluid Fermi gases \cite{Shin2008}. Based on previous studies, it is indeed very unlikely that phase separation could occur in our system. Although the surviving fermion density at long time is found to be below 15$\%$ of the initial one, \textit{both} spin states populations remain symmetrically populated. 
Moreover, even in the case in which the surviving fermions were almost spin polarized owing to the imperfect $2 \rightarrow 3$ transfer by the initial pump pulse at strong coupling, a complete phase separation of atoms and pairs would be surprising at such low fermion concentration \cite{Shin2008, Pilati2008}. 
Finally, it is important also to stress that phase segregation has been observed for \textit{degenerate} molecular gases well below the critical temperature $T_c$ for Bose Einstein condensation, whereas our paired component is found always at $T/T_c>1$.
    
\begin{figure}[t!]
\centering
\includegraphics[width= 8.7 cm]{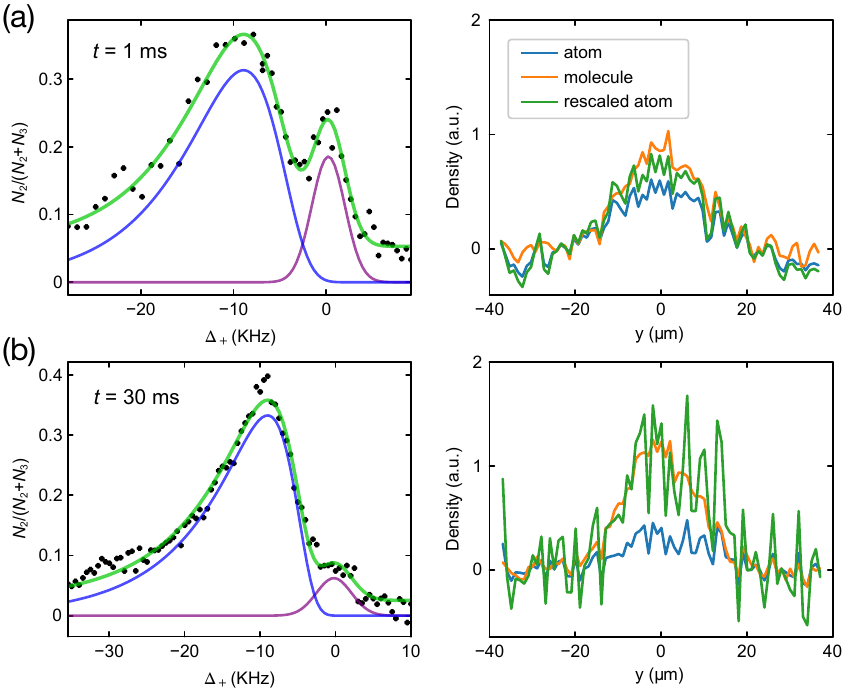}
\caption{\label{fig_atom_mol_profiles} 
Spectral response (left column) and in-situ radially integrated density profiles of the atomic and molecular populations (right column). The data shown are recorded after an evolution time of 1\,ms (\textbf{a}) and 30\,ms (\textbf{b}) from the repulsion quench via the RF pump pulse, with an interaction strength $\kappa_F a \simeq 2.3$. The experimentally obtained spectra are fitted using the sum of Gumbel (blue curve) and a Gaussian (purple curve) function, which account for the molecular and atomic response, respectively. The fitted functions are shown separately after zeroing their offsets for clarity.
}
\vspace*{-10pt}
\end{figure}

These expectations are indeed confirmed by the analysis presented in the following (see Fig.~\ref{fig_atom_mol_profiles}).
Figure ~\ref{fig_atom_mol_profiles} shows the spectral response (left column) and the radially-integrated in-situ profiles (right column) obtained for a system quenched at $\kappa_Fa \simeq 1.5$ and evolved in the interacting regime for 1\,ms (see Fig.~\ref{fig_atom_mol_profiles}(a)) and 30 ms (see Fig.~\ref{fig_atom_mol_profiles}(b)), respectively. 
To obtain selective information about the molecular (atomic) densities, for each time $t$ we acquired two distinct sets of images selectively probing the state-3 and state-2 population, following a fast and strong RF pulse detuned by $\Delta_+(t)$ from the free atom transition $\nu_0$. 
The optimized RF pulse transfers the remaining atoms in state 2, while leaving the molecules in state 3. 
To increase the signal to noise ratio, for both dimers and atoms we average 10 images acquired under the same experimental conditions, and we extract the profiles of the molecular and atomic densities by integrating the images along the radial direction.

As one can notice, the two density profiles are found always to macroscopically overlap, irrespective of the short or long evolution time, which only causes a change in the relative amplitudes of the two density distributions. 
For clarity, in the right panels of the figure we also show the atomic profiles properly rescaled in order to match the amplitude of the molecular signal.
From a closer inspection of the density profiles through Gaussian fits, we only find a slow and moderate shrinking of the molecular cloud relative to the initial atomic distribution. This is consistent with the fact that dimers experience a tighter harmonic confinement, twice as large as the atom one owing to the molecule polarizability two times higher than the one of single fermions.

The molecular distribution in the trap is found to reach a steady configuration after about 10-15 ms, a time which approximately corresponds to one quarter of the axial trap period.
At long evolution times, $t\geq 20$ ms, Gaussian fits to the low-density wings of the molecular and atomic distributions appear consistent with the reach of thermal equilibrium among the two components.
For $\kappa_F a \sim 2$ ($\kappa_F a \sim 1$), the estimated temperature after the equilibration time is about four (two) times larger than the initial one.
Notwithstanding the increased temperature and reduced overlap between the atomic and molecular components in the outer regions of the trap, the overall density near the trap center, dominated at long evolution times by the molecular phase, is equal or even slightly higher than the initial one $\overline{n}_0$.    

Besides the lack of any detectable mismatch in the two density distributions, Figure \ref{fig_atom_mol_profiles} also highlights how the direct detection of small clusters from the images is extremely challenging, despite our 1.3 $\mu$m imaging resolution, owing to line-of-sight integration throughout the inhomogeneous density distribution. As discussed in the main text and also in Section 1, information about the spatial extent of the magnetic domains and their location in the trap are more conveniently extracted by the analysis of density fluctuations from the absorption images.

\section{Spin density fluctuation measurements}
In the following we describe the experimental procedures followed to obtain a measurement of spin density fluctuations from in-situ absorption images of state-3 atoms, and the main outcomes of such characterization. 
As mentioned in the main text, the analysis of spin density fluctuations enables to estimate the gas temperature during the initial few milliseconds of evolution at strong coupling, and, possibly, to reveal the emergence of spin-polarized fermion domains into the system. 
While the study of density fluctuations allows for absolute thermometry in a system with a known equation of state \cite{MeinekePhD}, their interpretation in our experiment, after the quench to strong repulsion, is non-trivial. Here, the spin density fluctuations may arise both from an increase in the gas temperature and from ferromagnetic domain formation \cite{Sanner2012}. Nevertheless, even though we can not discriminate between the two mechanisms, this study provides useful upper bounds both for the temperature and the maximum size of the spin-polarized domains.

\begin{figure}[!t]
\vspace*{-3pt}
\includegraphics[width= 8.6 cm]{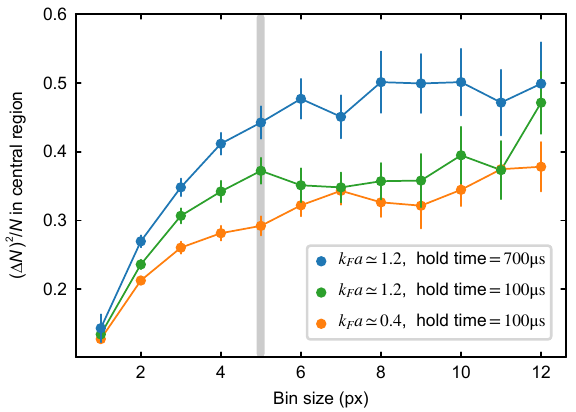}
\caption{Normalized atom number variance calculated in a central region of the cloud, corresponding to relative column densities of at least $0.7$ times  the peak density. The plotted curves correspond to the different interactions strength and hold times indicated in the legend. A bin size of 5\,px was used for all measurements presented Fig.4A-B of the main text, giving the best trade-off between a small bin size and a small blurring due to finite optical resolution, with $(\Delta N^2)$ close to saturation. 
}\label{fig:binning_size}
\vspace*{3pt}
\end{figure}

We extract in-situ spin density fluctuations by performing a statistical analysis over many experimental realizations for each chosen interaction strength and evolution time, recorded through state-resolved in-situ absorption imaging. Repulsive interactions are quenched in the gas using the RF pump procedure described in Section~\ref{S1C}. After a variable evolution time, we acquire $\sim 90$ absorption images of the state-3 cloud, prepared under identical experimental conditions, by means of a $5 \mu$s-long resonant pulse. The atom number variance $(\Delta N)^2 \equiv \text{Var}(N)$ is computed through the following procedure. First, we apply to each experimental optical density image a binning routine to bin the original pixels into larger super-pixels. 
Image binning is essential to avoid an underestimation of density fluctuations due to blurring, that results from the finite optical resolution of the imaging system \cite{Sanner2010, Muller2010}: since our imaging resolution $\sim 1.3\,\mu$m is comparable to the image pixel size $\text{px} \simeq 1.04\,\mu$m, it causes fluctuations in the content of a given column pixel to spread over the neighbouring pixels (see also the next Section). The effect of pixel binning on the final extracted $(\Delta N)^2$ is illustrated in Fig.~\ref{fig:binning_size}, where the atom number variance averaged over a central region of the cloud is plotted as a function of the bin size. Bins with a transverse area of $5 \times 5$\,px$^2$ are used for the data presented in Fig.~\ref{fig:136_raw} and Fig.~\ref{Fig_var1}.
Secondly, in order to suppress the effect of experimental shot-to-shot total atom number fluctuations on $(\Delta N)^2$, we automatically eliminate outliers by discarding the 10 experimental shots with highest and lowest total atom number. We then fit each optical density profile with a two-dimensional Gaussian envelope, that is subtracted from the image.  
\begin{figure}[!t]
\includegraphics[width= 8.6 cm]{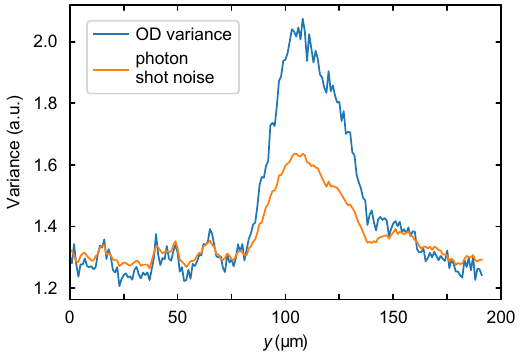}
\caption{A typical measurement of the optical density variance $(\Delta\,OD)^2$ (light blue) and the estimated photon shot noise $(\Delta_{\text{ph}}\,OD)^2$ (orange). A slice along the $y$-axis is shown. The CCD camera gain $G$ is obtained as a fit parameter by minimizing the difference between $(\Delta_{\text{ph}}\,OD)^2$ and $(\Delta\,OD)^2 $ in regions where no atoms are detected.}\label{fig:photonic_var}
\vspace*{-10pt}
\end{figure}

At this point, we compute the optical density variance $(\Delta \,OD)^2$ over the resulting data set containing $\mathcal{N} \approx 80$ images, to which various noise sources may contribute \cite{Sanner2010}. For the used imaging light and CCD sensor parameters, the measured $(\Delta \,OD)^2$ is largely dominated by two sources of fluctuations: the photonic shot noise of the imaging laser beam and the atomic density fluctuations. 
As we are only interested in atomic density fluctuations, i.e. the number fluctuations of the content of each binned volume, we remove the photonic shot-noise contribution $(\Delta_{\text{ph}} \,OD)^2$ from $(\Delta \,OD)^2$. By exploiting the Poissonian statistics of photon shot noise, we estimate $(\Delta_{\text{ph}}\,OD)^2$ through the mean photon counts in each pixel of the CCD sensor \cite{Sanner2010}:
$$(\Delta_{\text{ph}}\,OD)^2 = {1 \over G}\left({1 \over \langle I_\text{atoms} \rangle} + {1 \over \langle I_\text{ref} \rangle}\right),$$ 
where $G$ is the CCD sensor gain, and $\langle I_\text{atoms}\rangle$ and $\langle I_\text{ref}\rangle$ are the mean light intensity profiles recorded in the presence and absence of the atomic cloud, respectively, with their background levels subtracted.
Examples of $(\Delta\,OD)^2$ and $(\Delta_{\text{ph}}\,OD)^2$ obtained from a typical measurement run are shown in Fig.~\ref{fig:photonic_var}. The atomic noise contribution is obtained by subtracting the two: $(\Delta_{\text{atoms}} \,OD)^2 = (\Delta\,OD)^2 - (\Delta_{\text{ph}}\,OD)^2$.
Finally, the atom number variance $(\Delta N)^2$ in each super-pixel column of transverse area $A$ is obtained as $(\Delta N)^2 = (A/\sigma_{0c})^2 \, (\Delta_{\text{atoms}}\,OD)^2$, where $A \simeq 27\,\mu$m$^2$ and $\sigma_{0c}$ is the absorption optical cross section, which is independently calibrated. This calibration was found to be consistent with the expectation from a measurement of thermal density fluctuations of a degenerate ideal Fermi gas (see below). We have checked the dependence of the measured atom number variance upon the number of images for a given experimental realization, finding that 80 images are typically sufficient for the variance to stabilize.

The extracted $(\Delta N)^2$ is suitably normalized to the mean atom number $N \equiv \langle N \rangle$. In Fig.~\ref{fig:136_raw}, we show the variance of the atom number contained in each column bin as a function of the mean atom number in the same bin, for a weakly interacting Fermi gas at $\kappa_Fa \simeq 0.35$. Such measurement directly reveals the sub-Possonian nature of its density fluctuations due to Pauli blocking. 
For a degenerate ideal Fermi gas, density fluctuations are expected to be essentially the outcome of thermal fluctuations, with the two quantities connected by the fluctuation-dissipation theorem \cite{Muller2010, MeinekePhD, Sanner2010}. For the sub-volume $V_{ij}$ associated to each column bin, the temperature-induced atom number fluctuations are given by \cite{Muller2010, MeinekePhD}:
\begin{align}
{(\Delta N)_{ij}^2 \over N_{ij}} = {\text{Li}_2(\zeta_{ij}) \over \text{Li}_1(\zeta_{ij})}\,,
\label{eq: thermal_var}
\end{align}
where $\zeta_{ij}=e^{\beta \mu_{ij}}$ is the fugacity of the sample contained within the sub-volume $V_{ij}$, which is assumed to be in thermodynamic equilibrium with the rest of the gas.
By numerically inverting Eq.~\eqref{eq: thermal_var}, from the data shown in Fig.~\ref{fig:136_raw}, we obtain an estimate of the gas temperature of $T/T_F = 0.15(5)$. This value is consistent with $T/T_F = 0.12(2)$, that is independently determined by fitting the profile of the gas after a 5\,ms-long ballistic expansion with a Thomas-Fermi distribution. Such noise thermometry validates our procedure for extracting the atom number variance, and in particular the experimentally calibrated value of the optical cross section $\sigma_{0c}$. In order to further validate our procedure, we have also used computer-generated images of a thermal gas with atomic and photonic Poissonian noise, taking into account both the finite imaging resolution and the typical experimental shot-to-shot atom number fluctuations. In this case, we have verified that the extracted ${(\Delta N)^2 / N} \simeq 1$ across the entire image.

\begin{figure}[t!]
\includegraphics[width = 8.6 cm]{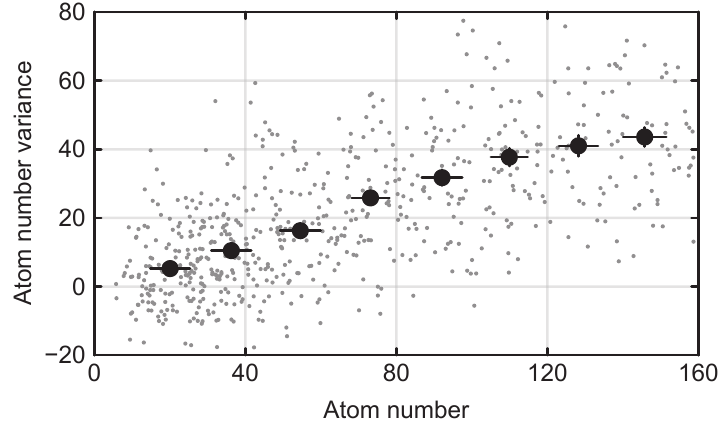}
\caption{Variance of a weakly interacting Fermi gas ($\kappa_Fa \simeq 0.3$) at zero evolution time. Small gray dots represent the atom number variance of each single super-pixels, with a bin size of 5\,px. The black circles result from further data binning in terms of average atom number, with error bars denoting the standard deviation of the mean of all data points within a single bin.}\label{fig:136_raw}
\end{figure}

\begin{figure}[!t]
\vspace*{5pt}
\includegraphics[width = 8.8cm]{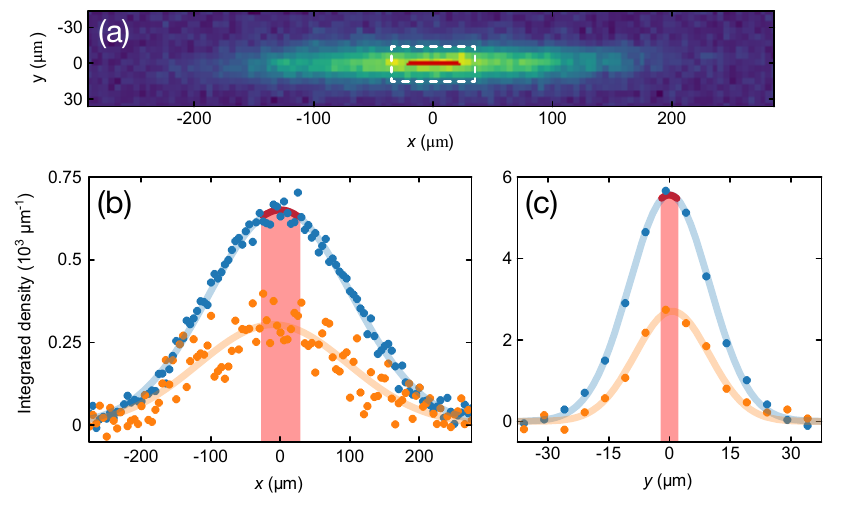}
\caption{Spatial dependence of the atom number variance for $t= 700\,\mu$s at $\kappa_F(0) \simeq 1.2$. (\textbf{a}) The column-integrated density profile resulting from averaging $\sim 80$ experimental images is shown, with the central analysis region $\Re$ marked in red.
The integration region used for spectroscopic probing (see Section S.1) is also indicated for comparison as a dashed white rectangle.
The radially (\textbf{b}) and axially (\textbf{c}) integrated profiles of the atom number mean (blue) and variance (orange) are also shown, with the same region as in panel (a) denoted by a red shading.
}\label{fig_high_var_region}
\vspace*{-5pt}
\end{figure}

\begin{figure}[b!]
\includegraphics[width = 8.3 cm]{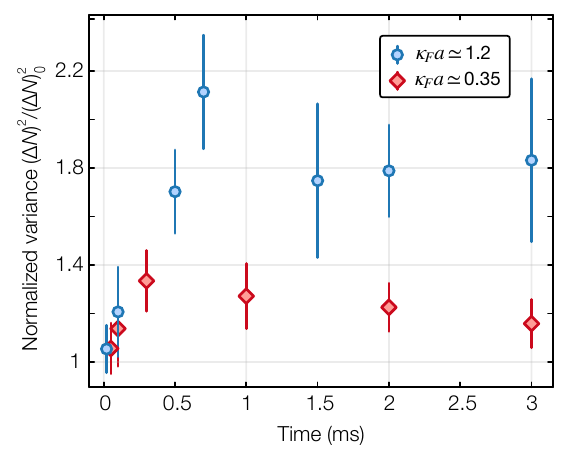}
\caption{Atom number variance $(\Delta N)^2$ normalized to that of an ideal Fermi gas $(\Delta N)_0^2$, measured within the central region $\Re$ at various times after the quench at $\kappa_F a \simeq 1.2$ (blue circles) and 0.35 (orange diamonds), respectively. Error bars denote the standard deviation of the mean.}\label{Fig_var1}
\end{figure}
%

%
We now turn to discuss the results of the real-time characterization of spin density fluctuations for different interaction strengths, acquired for evolution times up to 3\,ms after the quench, and focusing within the central region $\Re$ of the cloud where number of atoms per super-pixel is found within 10\% from the maximum one. 
In general, the growth of spin-polarized domains at strong coupling cannot be detected through direct spin-selective in-situ imaging of the cloud, owing to line-of-sight integration and to the small expected domain size $\xi$, only slightly larger than our imaging resolution of $\sim1.3\,\mu$m. 
Micro-scale spin inhomogeneities can however be revealed by the local measurement of spin density fluctuations \cite{Recati2010, Sanner2011, Sanner2012, Meineke2012}, since spin-polarized clusters lead to an increased variance of the spin number fluctuations inside a (column) probe volume with a transverse size comparable or larger than the imaging resolution \cite{Sanner2012}. 
In the simplistic case of Poissonian fluctuations, the enhancement of spin variance in a clusterized sample relative to that of a paramagnetic one would directly reflect the mean number of spins per domain \cite{Sanner2012}. On the other hand, an enhancement of spin variance in a non-clusterized sample, assumed to be locally in thermal equilibrium at each instant, can be attributed to an increased temperature of the gas. 

Figure~\ref{fig_high_var_region} shows the mean state-3 atom density distribution recorded after 700\,$\mu$s of evolution at $\kappa_Fa \simeq 1.2$. In Fig.~\ref{fig_high_var_region}(b)-(c) show the atom number  mean $\langle N \rangle$ and variance $(\Delta N)^2$ profiles integrated along the $y$- and $x$-axis. The region of the cloud marked in red indicates the typical analysis region $\Re$.
Figure~\ref{Fig_var1} presents the evolution of $(\Delta N)^2$ normalized to that of a non-interacting gas $(\Delta N)_0^2$, for interaction quenches at $\kappa_F a \simeq 0.35$ and 1.2, respectively. 
While for weak interactions, we observe only a small increase of fluctuations, compatible with a moderate enhancement of the gas spin susceptibility \cite{Pilati2010, Recati2010}, for $\kappa_F a \simeq 1.2$ we detect a more pronounced growth of $(\Delta N)^2$. 
The observed maximum two-fold increase of the column-integrated spin fluctuations $(\Delta N)^2$ allow for two distinct interpretation: on the one hand, it seems compatible with spin separation occurring only on a small $\mu$m-scale of very few interparticle spacings, in agreement with the domain-size estimate based on our measurement of $\Gamma_{\Delta}$, i.e. $\xi \approx 2\pi/\kappa_F \sim 2.5\,\mu$m \cite{Pekker2011}.
On the other hand, the observed trend at strong coupling can also be interpreted as a consequence of local heating of the gas, rapidly evolving under the concurrent action of ferromagnetic and pairing instabilities at short times. 
In this perspective, our data suggest a rapid two-fold (local) temperature increase established already at 1\,ms after the quench, the atom-number fluctuations always remaining sub-Poissonian \cite{Sanner2010, Muller2010}, owing to Pauli blocking in degenerate Fermi gases. 
The system temperature obtained via such an analysis appears consistent with the one obtained at longer evolution times $t \geq 10$ ms, estimated through Gaussian fits to the wings of the in-situ density distribution of atomic and molecular components at the same $\kappa_F a$.

\end{document}